\documentclass[aps,prd, nofootinbib,preprintnumbers]{revtex4}%
\usepackage{subcaption}
\usepackage{amsmath,amsfonts,amssymb,graphicx,graphics,color, hyperref}
\usepackage[latin9]{inputenc}
\usepackage{pifont}
\usepackage{natbib}
\usepackage{epsfig,epstopdf}
\usepackage{bm}
\usepackage{enumerate}
\usepackage{graphicx}
\usepackage{amsmath}
\usepackage{amsfonts}
\usepackage{amssymb}
\usepackage[export]{adjustbox}
\usepackage{booktabs}
\usepackage{svg}%
\providecommand{\U}[1]{\protect\rule{.1in}{.1in}}

\newcommand{\be}{\begin{equation}}
\newcommand{\ee}{\end{equation}}
\newcommand{\bea}{\begin{eqnarray}}
\newcommand{\eea}{\end{eqnarray}}

\begin{document}
\title{Detecting the finer structure of the P vs NP problem with statistical
mechanics: the case of the Wang tiling problem}
\author{Fabrizio Canfora$^{1,2}$}
\author{Marco Cedeño$^{2}$ }
\affiliation{$^{1}$Centro de Estudios Científicos (CECs), Avenida Arturo Prat 514,
Valdivia, Chile.}
\affiliation{$^{2}$Facultad de Ingeniería, Universidad San Sebastian, sede Valdivia,
General Lagos 1163, Valdivia 5110693, Chile.}
\email{fabrizio.canfora@uss.cl, marco.cedeno@uss.cl }

\begin{abstract}
We introduce the idea that the P vs NP problem can have a \textit{finer
structure}. Given the NP complete problem of interest, the configurations
space of the problem can be divided in (at least) two regions. In one region,
polynomial algorithms to solve the NP complete problem of interest are
available (and we discuss one possible realization inspire by the games of
chess and go). In the second region the problem to find polynomial time
algorithms is very similar to the problem to find polynomial time algorithms
to determine the asymptotic behavior of discrete dynamical systems in the
chaotic regime. We cannot exclude the existence of a third region which
separates the first two: this region would have the characteristics of "the
edge of chaos". We focuss on the Wang tiling problem of an $N\times N$ square
(with $N$ large): here a Wang tiles set $\Gamma$ is an "alphabet". We
construct a
\"{}%
statistical-physics inspired%
\"{}
heuristic which allows to define \textit{good alphabets} as the ones with a
good thermodynamical behavior. For (a suitable subclass of) \textit{good
alphabets} we construct an algortihm which, in polynomial time, determines how
to tile the $N\times N$ square. On the other hand, for bad alphabets, we
observe a
\"{}%
chaotic behavior%
\"{}%
. The Cook-Levin theorem advocate a similar pattern for all the NP-complete problems.

\end{abstract}
\maketitle
\tableofcontents

\section{Introduction}

The \textbf{P versus NP problem} was introduced by Stephen Cook and Leonid
Levin (see \cite{PvsNP1} \cite{PvsNP2}: for detailed review see \cite{PvsNP3}
\cite{PvsNP5} and references therein). On the other hand, both John Nash and
Kurt Gödel had a clear intuition of the core of the problem (see
\cite{PvsNP3.5} and \cite{PvsNP4}) before the pioneering results by Cook and
Levin. Roughly speaking, the \textbf{P versus NP problem} (henceforth P vs NP)
asks whether or not every problem whose solution can be verified in a
polynomial time can also be solved in polynomial time. Although the P versus
NP was officially born in 1971, the discussions on the difficulty of proof as
well as on the potential consequences started well before. In 1955 J. Nash was
analyzing the question of whether or not the time required to crack a complex
code would increase exponentially with the length of the key. Kurt Gödel (in a
letter to John von Neumann in 1956) argued that the discovery of mathematical
proofs could be automated provided one could solve in quadratic (polynomial)
time the "theorem-proving" (see the references here above).

There is no doubt that the \textbf{P versus NP problem} is one of the most
important open problems not only in mathematics and computer science but,
actually, in science in general. The consequences to solve it would be huge
(see, for instance, \cite{PvsNP3} \cite{PvsNP5} and references therein).

The implications of resolving the P versus NP dilemma extend far beyond the
boundaries of theoretical computer science, permeating fields such as
cryptography, optimization, and statistical physics. A potential collapse of
NP into P would fundamentally alter our understanding of algorithmic
efficiency, rendering modern public-key cryptography insecure while
potentially revolutionizing fields reliant on complex optimization
\cite{PvsNP3}. From a physical perspective, the intractability of NP-complete
problems is deeply intertwined with the thermodynamics of disordered systems,
where the search for a ground state in spin glasses provides a direct physical
analogue to minimizing cost functions in combinatorial optimization
\cite{Kirkpatrick1983}. Seminal work at this intersection has revealed that
the computational hardness often arises from phase transitions in the solution
space, a phenomenon where statistical mechanics provides powerful tools for
analyzing complexity thresholds \cite{Monasson1999}. Consequently, exploring
heuristic strategies inspired by physical dynamics and solution-space geometry
remains a critical avenue for tackling these computationally hard barriers,
offering new insights into problems traditionally viewed solely through an
algorithmic lens \cite{Mezard2002}.

Despite these historical successes, current challenges lie in overcoming the
algorithmic barriers imposed by the rugged energy landscapes of hard
optimization problems. Recent approaches have shifted towards integrating
statistical physics with deep learning, specifically through physics-inspired
Graph Neural Networks (GNNs) capable of solving combinatorial problems by
mimicking relaxation processes in disordered media \cite{Schuetz2022}.
However, theoretical constraints persist; recent rigorous results on the
Overlap Gap Property (OGP) suggest that for random instances of NP-hard
problems, the solution space decomposes into disconnected clusters, creating
an algorithmic hardness that essentially blocks even the most advanced local
algorithms from finding optimal solutions in polynomial time
\cite{Gamarnik2021}. Consequently, the frontier of research focuses on
understanding these topological barriers to design "landscape-aware"
algorithms that can navigate or bypass these frozen states
\cite{Zdeborova2020}.

In the present paper we discuss the finer structure of this problem. Namely,
in the present manuscript, we provide with sound basis the following
intriguing possibility:

\textit{Within the configurations spaces of NP-complete problems}, specific
topological signatures exist that allow for efficient navigability.

To ground this hypothesis, we must revisit the rigorous definition of the
class at the heart of this inquiry. A decision problem $\mathcal{L}$ is
classified as \textbf{NP-complete} if it satisfies two conditions: first,
$\mathcal{L}$ must belong to the class NP (verifiable in polynomial time), and
second, every other problem in NP must be polynomial-time reducible to
$\mathcal{L}$ \cite{Karp1972}. This universality implies that an efficient
algorithm for any single NP-complete problem --such as the tiling of the plane
or Boolean
\"{}%
satisfability%
\"{}%
-- would automatically yield efficient solutions for the entire class,
bridging vast disciplines from operations research to quantum many-body
physics \cite{GareyJohnson1979, Preskill2018}.

Based on this definition, we postulate that the solution space is not
homogeneous. Instead, we propose the existence of a structural bifurcation
dividing the configuration space into (at least) two distinct regimes:

\begin{enumerate}
\item \textbf{The Tractable Regime:} In this first sub-region, the topological
structure allows for the construction of polynomial-time algorithms. We will
introduce a specific construction (inspired by the games of \textit{chess} and
\textit{go}) demonstrating this possibility later in the manuscript.

\item \textbf{The Chaotically Hard Regime:} This second sub-region is
\textit{genuinely computationally hard}. Here, the absence of efficient
algorithms is not merely a lack of knowledge but likely an intrinsic property
of the system. The probability of finding a polynomial-time algorithm to solve
instances in this region is analogous to the likelihood of finding a shortcut
to predict the asymptotic behavior of a discrete dynamical system in its
chaotic phase---a phenomenon often referred to as computational irreducibility
\cite{Wolfram1984}.

\item \textbf{Edge of chaos regime}: we cannot exclude the existence of this
third region which does satisfy neither the criteria to be in the tractable
regime nor in the chaotic regime. This sub-region would play the role of a
boundary separating the first two.
\end{enumerate}

We will focus the present analysis on the bounded version of the Wang tilings
problem\footnote{A Wang tile is a square with one color for each side. Given a
finite family ("alphabet") of different Wang tiles (from now on, a Wang tile
will simply be denoted as tile), the problem is to understand whether or not
one can cover the plane using these tiles and satisfying certain rules (to be
introduced in the next sections: see \cite{[1]}).} \cite{[1]} (which is the
prototype of NP-complete problem in computer science with deep connections to
statistical physics). The undecidable nature of the Wang tilings problem (WTP
henceforth) for arbitrarily large squares is closely related to the existence
of aperiodic tile sets (as if all the tilings would be periodic then the WTP
would be decidable). In fact, the WTP is undecidable \cite{[2]} and aperiodic
tilings have been found (moreover, to construct aperiodic Wang tiles with
small alphabets is a very active field in itself). As it is by now well known
(see for instance \cite{UndecidPvsNP}\ and references therein), the
undecidable nature of the WTP in unbounded domains is closely related to the
fact that the bounded version of WTP is NP-complete.

The boundary \textit{in the space of all alphabets} which we will discuss in
the following sections separates alphabets for which the decision problem can
be solved in a polynomial time from the ones for which it is very unlikely to
find a polynomial algorithm. Such boundary is closely related to the
transition to chaos in discrete dynamical systems. In view of
\cite{UndecidPvsNP}, one can also see that such a boundary separates the space
of alphabets into a decidable region and a genuinely undecidable region. Our
analysis does not exclude the existence of a sort of "grey region" similar to
the bifurcation cascades characterizing the transition to chaos in discrete
dynamical systems (for a review see \cite{[15]}). Namely, it is possible that
the boundary between subregions 1 and 2 is actually a third subregion
characterized by bifurcation cascades (which is a typical route to chaos in
discrete dynamical systems).

The idea of our framework \cite{MarcoFabrizio}, inspired by statistical
mechanics, is to define suitable temperature, entropy and partition function
associated to any alphabet $\Gamma$ in order to define an effective heuristic
able to identify a subset of alphabets for which the WTP can be solved in a
polynomial time. The key observation is that if such auxiliary statistical
mechanical system (associated to a given alphabet $\Gamma$) does possess a
good thermodynamic limit (in which case $\Gamma$ is a good alphabet) then the
corresponding alphabet $\Gamma$ is a good candidate to tile the
plane\footnote{Indeed, even if alphabets which satisfy our criteria of "being
good" for a large number of steps but which fail to tile the whole $%
\mathbb{R}
^{2}$ exist, it is very unlikely to meet these in practical applications. In
particular, in \cite{MarcoFabrizio} it has been shown that all the available
alphabets in the literature which satisfy the criteria to be good alphabets
(for a not so large number of steps) do, in fact, tile the whole $%
\mathbb{R}
^{2}$.} and, moreover, the number of tilings grows very fast as the size of
the squares increases. This fact will allow us to devise a strategy to find
polynomial-time algorithms for good alphabets inspired by the games of chess
and go (in the sense that one can sacrifice something at the beginning to get
greater benefits in the future). These algorithms work in a subset of the
\textit{good alphabets}. However we will discuss that this could be due to the
limitations impose by our available hardware and that the proposed algorithms
could work in a very large part of the \textit{good alphabets}. Thus, in order
to construct polynomial time algorithms for good alphabets, a key role will be
played by statistical mechanics.

At this point, the following comment is in order: when a given alphabet is not
good (according to our definition), it does not necessarily mean that there is
no polynomial-time algorithm. Instead, when the alphabet is not good, its
behavior (in a precise sense to be specified in the next sections) is chaotic
when one increases the size of the square to tile (something which prevents
one from guessing what happens in the case of very large squares). Thus, when
the alphabet of interest is not good, we can only conclude that the likelihood
to find polynomial-time algorithms is similar to the likelihood to find
polynomial-time algorithms to predict the asymptotic behavior of discrete
dynamical systems in the chaotic regime.

It is worth emphasizing here some pioneering results on the WTP. Robinson
\cite{[3]} improved the original results of Berger on the undecidability of
the WTP in the large constructing a smaller aperiodic tile sets. Along this
line, Jeandel and Rao \cite{[4]} constructed the smallest aperiodic tileset
(with eleven tiles). Despite the fact that in most of the examples of
aperiodic tiles, self-similar hierarchical structures appear (see \cite{[4]}
\cite{[4.1]} and references therein) the tiling constructed in \cite{[4.2]} is
neither self-similar nor periodic: this tiling is an important benchmark for
our protocols.

The WTP together with its characteristic aperiodic structures is related to
many problems in physics (see, for instance, \cite{[5]} \cite{[6]} \cite{[7]}
\cite{[8.1]} \cite{[9]} \cite{[10]} \cite{[11]}\ \cite{[11.1]} \cite{[13]}
\cite{[14]} and references therein). However, the deepest connection of WTP
with physics has to do with chaos theory. In our approach, it is possible to
associate a discrete mapping to any family of Wang tiles (namely, to any
\textit{Wang alphabet}) in such a way that when such mapping is chaotic, it is
unlikely to find a polynomial algorithm to tile large squares.

The Wang tiling problem is related to chaotic systems (for a review see
\cite{[15]}) and this is a particular case of the close connection between
undecidability and deterministic chaos (see \cite{15a0} \cite{15a} \cite{15d}
\cite{15e} and references therein). These \ results strongly suggests that
NP-completeness and undecidability manifest themselves in chaotic dynamical
systems (see \cite{UndecidPvsNP} and references therein).

From a computer science perspective, these results offer crucial insights into
the nature of intractability. The problem of determining whether a given
finite set of tiles can cover a simply connected region is known to be
NP-complete, implying that, assuming $P\neq NP$, no polynomial-time algorithm
exists for the general case \cite{Arora2009}. Specifically, \cite{Pak2013}
demonstrated that tiling simply connected regions with a fixed set of
rectangles remains NP-complete, establishing that computational hardness
persists even when the geometric constraints appear simplified. This inherent
complexity renders exhaustive search methods---which scale exponentially in
time and energy consumption---infeasible for practical applications.
Consequently, there is an urgent need to develop efficient criteria (or
heuristics), or \textquotedblleft recipes,\textquotedblright\ capable of
deciding \textquotedblleft tilability\textquotedblright\ by identifying
polynomial-time verifiable invariants rather than resorting to prohibitive
brute-force enumeration. This strategy aligns with recent breakthroughs such
as those by \cite{Aamand2023}, who successfully derived polynomial-time
algorithms for specific sub-classes of Polyominoes and domino packings.
Furthermore, the recent discovery of aperiodic monotiles (the so-called
\textquotedblleft Einstein\textquotedblright\ tiles) underscores the subtle
boundary between decidable patterns and complex, non-periodic order, renewing
interest in the topological constraints of tiling \cite{Smith2024}. It is
worth emphasizing that our proposed approach is designed to be robust,
allowing for extensions to diverse tiling variants and potentially to
higher-dimensional spaces.

Our work builds upon these foundations by proposing "physically inspired"
heuristics serving as certificates of "tilability" for tiles alphabets. By
verifying these heuristics, one can preemptively determine the feasibility of
tiling a region, thus avoiding unnecessary computational overhead. This not
only contributes to the theoretical understanding of "tilability" within the P
versus NP landscape but also offers huge practical benefits in applications
where resource constraints are critical.

This paper is organized as follows: in the second section we introduce the
Wang tiling problem together with the "decidability" protocols. In the third
section we discuss the dynamical characterization of the selected good
alphabets. In the fourth section, we will discuss the relations of the present
approach with discrete chaos. In the fifth section, we will discuss the
structure of the
\"{}%
Tiling Space%
\"{}%
. In the sixth section, we introduce the polynomial time algorithms for good
alphabets. In the seventh section we analyze the computational performance of
the algorithm and its scalability. In the final section some conclusions will
be drawn.

\section{Review of the Wang tiling problems and the $W_{\Gamma}(n)$ mapping}

Here we will shortly describe the WTP together with the heuristic introduced
in \cite{MarcoFabrizio} to distinguish good from bad alphabets. A Wang tiles
set $\Gamma$ or "alphabet" is a collections of a fixed number of $q$ different
square tiles (the size of the alphabet will be denoted as $q=\left\vert
\Gamma\right\vert $). Each edge of the tiles possesses a color (a discrete
label). The matching rules are the following:

\textbf{1) }The squares can be neither rotated nor reflected

\textbf{2) }Two square can be matched side to side only if the right edge of
the left square has the same color as the left edge of the right square.

\textbf{3) }Two square can be matched top to bottom only if the bottom edge of
the square on the top has the same color as the top edge of the square on the bottom.

Then, we want to guess whether or not a given alphabet $\Gamma$ of size $q$ is
a good candidate to cover up the whole plane. The idea is to find a concrete
boundary in the space of all alphabets $\Gamma$ separating the good candidates
from the rest. The results in \cite{MarcoFabrizio} show that in the space of
all alphabets such a boundary can be defined. Furthermore, it shares many
features with the boundary separating regular from chaotic dynamics in
discrete dynamical systems.

Let us define the following quantity:
\begin{equation}
W_{\Gamma}\left(  n\right)  \overset{def}{=}\ \left\{
number\ of\ different\ tilings\ of\ a\ n\times
n\ squares\ with\ the\ alphabet\ \Gamma\right\}  \ . \label{map1}%
\end{equation}
Let us also define \textit{the energy} $E_{\Gamma}(n)$ \textit{of an alphabet}
$\Gamma$ on a $n\times n\ square$ as the area $A(n)$ of the square (namely
$n^{2}$):%
\begin{equation}
E_{\Gamma}(n)\overset{def}{=}A(n)=n^{2}\ . \label{map1.5}%
\end{equation}
Hence $W_{\Gamma}\left(  n\right)  $ \textit{plays the role of the degeneracy
of the energy level} $E_{\Gamma}(n)$. Consequently, \textit{the
entropy\footnote{In the following it will be convenient to use the $\log_{10}$
instead of the usual $\ln.$}} $S_{\Gamma}(n)$ of $\Gamma$ \textit{at the
energy level} $n$ is
\begin{equation}
S_{\Gamma}(n)\overset{def}{=}\log W_{\Gamma}\left(  n\right)  \ .
\label{map1.75}%
\end{equation}

We expect that for good alphabets $\Gamma$ (with have a normal thermodynamical
behavior) there will be many different ways to tile a $n\times n$ square
(since for most of physically reasonable systems the degeneracy of the energy
level is an increasing function of the energy). In these cases $W_{\Gamma
}\left(  n\right)  $ is different from zero for arbitrarily large $n$ (which
is a rather obvious fact in many physical systems). From the WTP viewpoint
this precisely means that $\Gamma$ can tile arbitrarily large regions.
Therefore for alphabets with a normal thermodynamical behavior (which we
define as \textit{good alphabets}) both $W_{\Gamma}\left(  n\right)  $ and
$S_{\Gamma}(n)$ are be increasing function of $n$ and, consequently, good
alphabets can tile arbitrarily large regions. This is equivalent to the usual
requirement that the temperature should be positive:
\begin{equation}
\frac{\partial S}{\partial E}>0\Leftrightarrow\ \frac{\partial S_{\Gamma}%
(n)}{\partial n}>0\ . \label{map1.9}%
\end{equation}

\textit{From statistical mechanics, }we know that the partition function
captures the main characteristics of a system. Hence we can define \textit{the
partition function of the alphabet} $\Gamma$ \textit{as} $Z_{\Gamma}$:%
\begin{equation}
Z_{\Gamma}(\beta)\overset{def}{=}\sum_{n}W_{\Gamma}\left(  n\right)
\exp\left(  -\beta n^{2}\right)  \ . \label{map1.99}%
\end{equation}
The above partition function encodes relevant informations about the alphabet
$\Gamma$ (see the discussion in \cite{MarcoFabrizio}).

\subsection{Protocol I}

In order to devise an effective heuristic to detect a subclass of good
alphabet, the requirement to satisfy the condition in Eq. (\ref{map1.9}) is a
good starting point. Since it is not possible to compute $W_{\Gamma}\left(
n\right)  $ for arbitrarily large $n$, one can only verify that the condition
in Eq. (\ref{map1.9}) is satisfied for finite $n$. Thus, the question is: how
large should $n_{\max}$ be in order to be confident that the $\Gamma$ is a
good candidate? The results in \cite{MarcoFabrizio} suggest that it could be
enough to verify Eq. (\ref{map1.9}) for all $n\leq n_{\max}$ with
\begin{equation}
n_{\max}\gg\left\vert \Gamma\right\vert =q. \label{protocol1a}%
\end{equation}
In the present manuscript, our computing power only allows to take $n_{\max}$
2 or 3 times $q$ (which is not enough to disclose the full power of our
proposal but it is good enough).

\subsection{Protocol II}

It is possible to define a second heuristic to detect the "goodness" of a
given alphabet $\Gamma$ disclosing at the same time a very intriguing relation
with chaos theory in discrete dynamical systems. The idea is to plot
$W_{\Gamma}\left(  n+1\right)  $ in terms of $W_{\Gamma}\left(  n\right)  $.
In this way, the mapping $f_{\Gamma}$ is computed numerically plotting on the
vertical $Y-$axis $W_{\Gamma}\left(  n+1\right)  $ and on the horizontal
$X-$axis $W_{\Gamma}\left(  n\right)  $:
\begin{equation}
P=(X,Y)=\left(  W_{\Gamma}\left(  n\right)  ,W_{\Gamma}\left(  n+1\right)
\right)  \ . \label{map2}%
\end{equation}
In this way we define $W_{\Gamma}\left(  n+1\right)  $ as a function
$F_{\Gamma}$ of $W_{\Gamma}\left(  n\right)  $:
\begin{equation}
W_{\Gamma}\left(  n+1\right)  =F_{\Gamma}\left(  W_{\Gamma}\left(  n\right)
\right)  \ . \label{map3}%
\end{equation}
Such function $F_{\Gamma}$ is the sought discrete mapping. We will do the same
with the entropy, plotting $S_{\Gamma}\left(  n+1\right)  $ in terms of
$S_{\Gamma}\left(  n\right)  $ obtaining
\begin{equation}
S_{\Gamma}\left(  n+1\right)  =f_{\Gamma}\left(  S_{\Gamma}\left(  n\right)
\right)  \ , \label{map3.1}%
\end{equation}
$S_{\Gamma}\left(  n+1\right)  $ as a function $f_{\Gamma}$ of $S_{\Gamma
}\left(  n\right)  $. The undecidable nature of WTP suggests that is
impossible to compute explicitly $f_{\Gamma}$ for a generic $\Gamma$. In fact
this is not necessary at all. The results in \cite{MarcoFabrizio} shows that
$F_{\Gamma}$ and $f_{\Gamma}$ behave very differently depending on whether
$\Gamma$ is good or bad. This fact allows to deduce many intriguing properties
despite the fact that the analytic form of $f_{\Gamma}$ is not known precisely.

\section{Dynamical Characterization of Selected ``Good'' Alphabets}

In order to deeply analyze the structural and dynamical properties of Wang
tiles, we employ a threefold graphical approach for each alphabet $\Gamma$.
The diagnostic framework consists of: (i) the geometric structure of the
tiling, (ii) the complexity growth behavior $W_{\Gamma}(n)$ as a function of
$n$ (essential for the first heuristic), and (iii) the phase-space recurrence
plot $W_{\Gamma}(n+1)$ versus $W_{\Gamma}(n)$, which informs the second
heuristic. While this methodology allows for detection of both efficient and
inefficient behaviors, in this section, it is utilized to reveal the stable
combinatorial features inherent to robust alphabets. For every analyzed case,
these three components are integrated into a single figure to facilitate a
direct visual correlation between topological structure and dynamical performance.

Adhering to the classification framework established by \cite{MarcoFabrizio},
we adopt the notation $G_{i}$ for \textquotedblleft good\textquotedblright%
\ alphabets and $B_{i}$ for \textquotedblleft bad\textquotedblright\ ones.
While previous studies have utilized the aforementioned plots to discriminate
between these two categories identifying specific instances of $B_{i}$
characterized by high entropy but poor tiling capabilities the present work
restricts its scope exclusively to the $G_{i}$ class. We assume the
distinction criteria as a premise, focusing instead on the specific growth
signatures and phase-space stability that define these effective alphabets.

Consequently, this analysis centers on a subset of five specific ``good''
alphabets, selected through a combination of empirical experimentation and
validation against established literature, as detailed in Table
\ref{good_families}. In the following subsections, a detailed description is
provided for each of these five families. The discussion is supported by the
tripartite visualization scheme, linking the physical tile set $\Gamma$ with
its complexity growth $W_{\Gamma}(n)$ and its phase-space trajectory
$W_{\Gamma}(n+1)$ vs $W_{\Gamma}(n)$, setting the stage for the subsequent
heuristic analysis of mapping growth.

\begin{table}[ptbh]
\centering
\begin{tabular}
[c]{cll}\hline
\toprule \textbf{Family ID} & \textbf{Source} & \textbf{Observation}\\\hline
\midrule G1 & Empirical experimentation & Two tiles self similar\\
\addlinespace G2 & B. Grünbaum \cite{[19]} & Aperiodic and self-similar tiling
system\\
\addlinespace G3 & E. Jeandel \cite{[4]} & Aperiodic and non-self-similar
tiling system\\
\addlinespace G4 & S. Labbé \cite{[4.1]} & Aperiodic and self-similar tiling
system\\
\addlinespace G5 & K. Culik \cite{[23]} & Aperiodic and non-self-similar
tiling system\\\hline
\bottomrule &  &
\end{tabular}
\caption{Selected alphabet and their sources.}%
\label{good_families}%
\end{table}

\begin{figure}[htbp]
	\centering
	\begin{subfigure}[c]{0.15\linewidth}
		\centering
		\includegraphics[width=\linewidth]{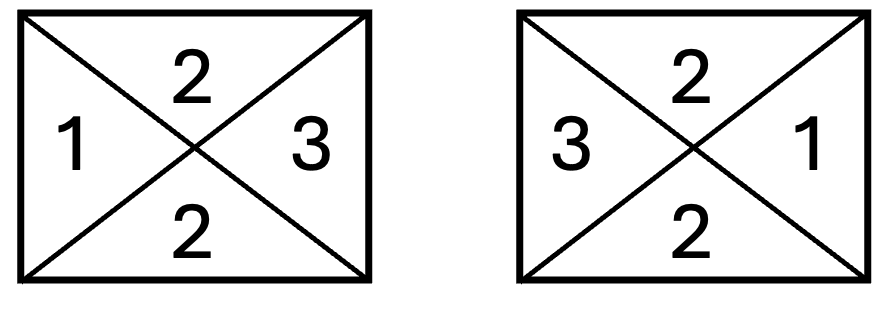}
		\caption{Alphabet G1}
		\label{fig:G1_analysis_a}
	\end{subfigure}
	\hfill 
	\begin{subfigure}[b]{0.4\linewidth}
		\centering
		\includegraphics[width=\linewidth]{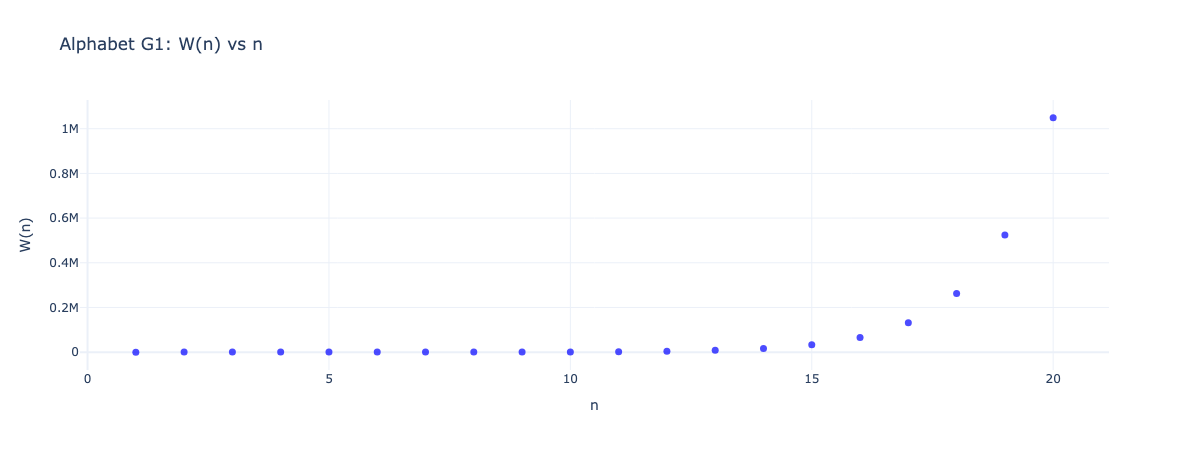}
		\caption{$W_{\Gamma}(n)$ vs $n$}
		\label{fig:G1_analysis_b}
	\end{subfigure}
	\hfill
	\begin{subfigure}[b]{0.4\linewidth}
		\centering
		\includegraphics[width=\linewidth]{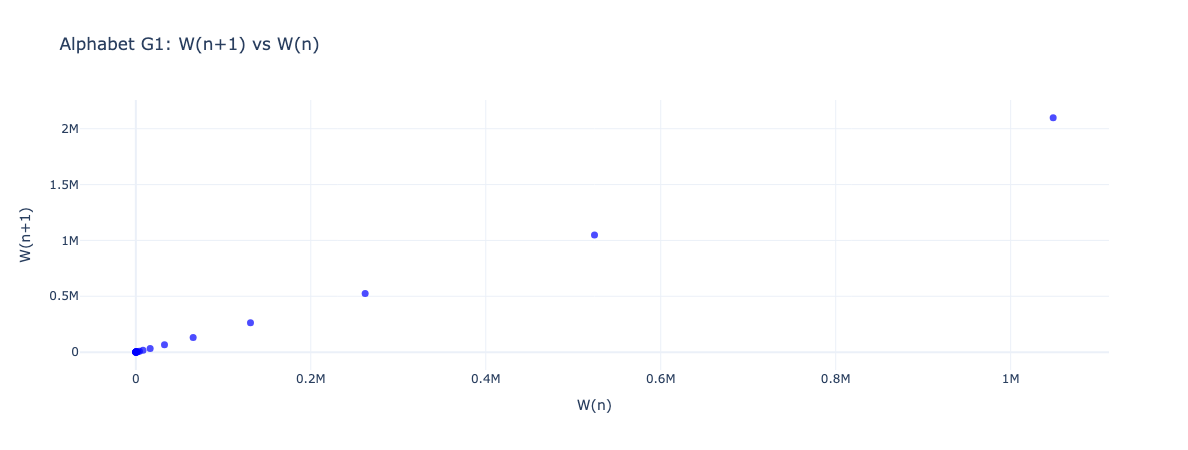}
		\caption{$W(n+1)$ vs $W(n)$}
		\label{fig:G1_analysis_c}
	\end{subfigure}
	
	\caption{Alphabet G1 and its associated complexity plots.}
	\label{fig:G1_analysis}
\end{figure}

\textbf{Alphabet G1.} This family represents the canonical baseline for a
functional or \textquotedblleft good\textquotedblright\ alphabet. Despite
comprising a minimal set of only two elements, it successfully achieves a
valid tessellation of the entire two-dimensional plane. The complexity
analysis, denoted by $W_{\Gamma}(n)$, reveals a strictly monotonic increase as
the system size $n$ scales. Corroborating this regularity, the phase-space
trajectory $W_{\Gamma}(n+1)$ versus $W_{\Gamma}(n)$ manifests a distinct
linear regime. This behavior underscores the structural simplicity and
deterministic coverage capacity of the set, as illustrated in
Fig.~\ref{fig:G1_analysis}.

\begin{figure}[htbp]
	\centering
	\begin{subfigure}[b]{0.2\linewidth}
		\centering
		\includegraphics[width=\linewidth]{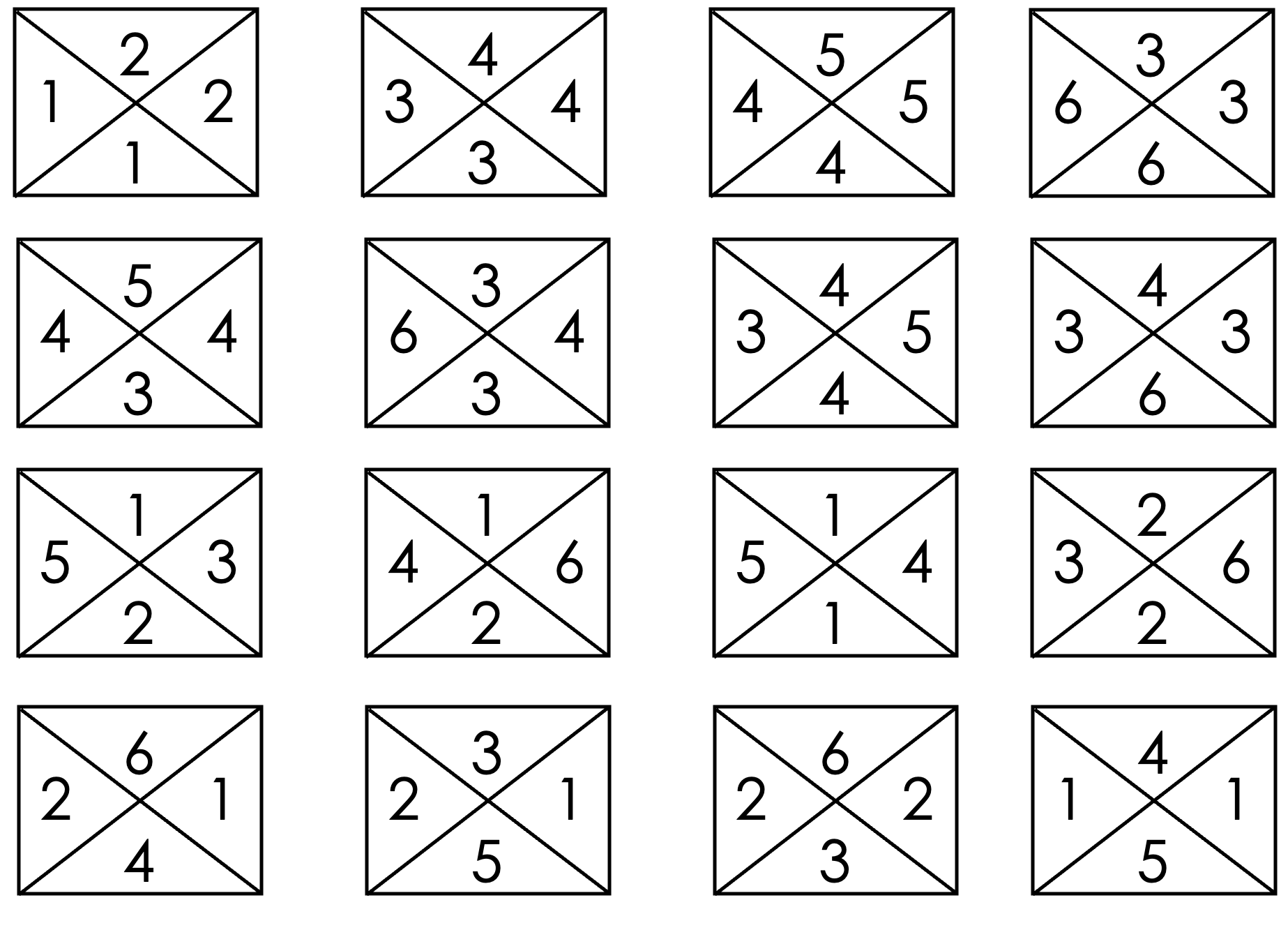}
		\caption{Alphabet G2}
		\label{fig:G2_analysis_a}
	\end{subfigure}
	\hfill 
	\begin{subfigure}[b]{0.32\linewidth}
		\centering
		\includegraphics[width=\linewidth]{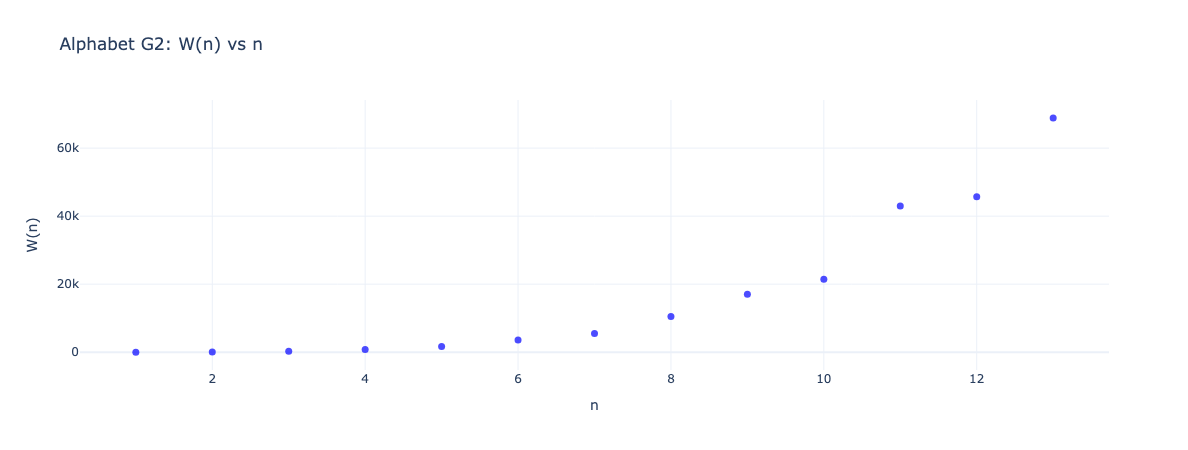}
		\caption{$W_{\Gamma}(n)$ vs $n$}
		\label{fig:G2_analysis_b}
	\end{subfigure}
	\hfill 
	\begin{subfigure}[b]{0.32\linewidth}
		\centering
		\includegraphics[width=\linewidth]{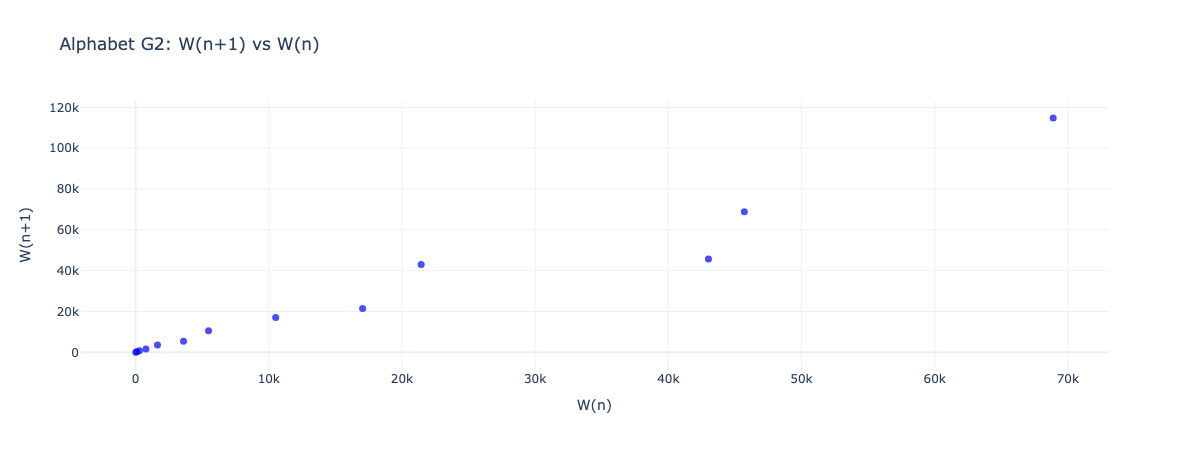}
		\caption{$W(n+1)$ vs $W(n)$}
		\label{fig:G2_analysis_c}
	\end{subfigure}
	
	\caption{Alphabet G2 and its associated complexity plots.}
	\label{fig:G2_analysis}
\end{figure}

\textbf{Alphabet G2.} Derived from the study by \cite{[19]}, this set serves
as a paradigmatic example of an aperiodic tiling system driven by
self-similarity. While the complexity curve $W_{\Gamma}(n)$ follows a
nonlinear trajectory, it maintains a stable monotonic ascent. Crucially, the
phase-space projection $W_{\Gamma}(n+1)$ versus $W_{\Gamma}(n)$ captures the
underlying hierarchical and recursive architecture of the assembly. This
intrinsic structural coherence is the defining attribute that validates its
categorization as a \textquotedblleft good\textquotedblright\ alphabet. The
corresponding graphical analysis is presented in Fig.~\ref{fig:G2_analysis}.

\begin{figure}[htbp]
	\centering
	\begin{subfigure}[b]{0.2\linewidth}
		\centering
		\includegraphics[width=\linewidth]{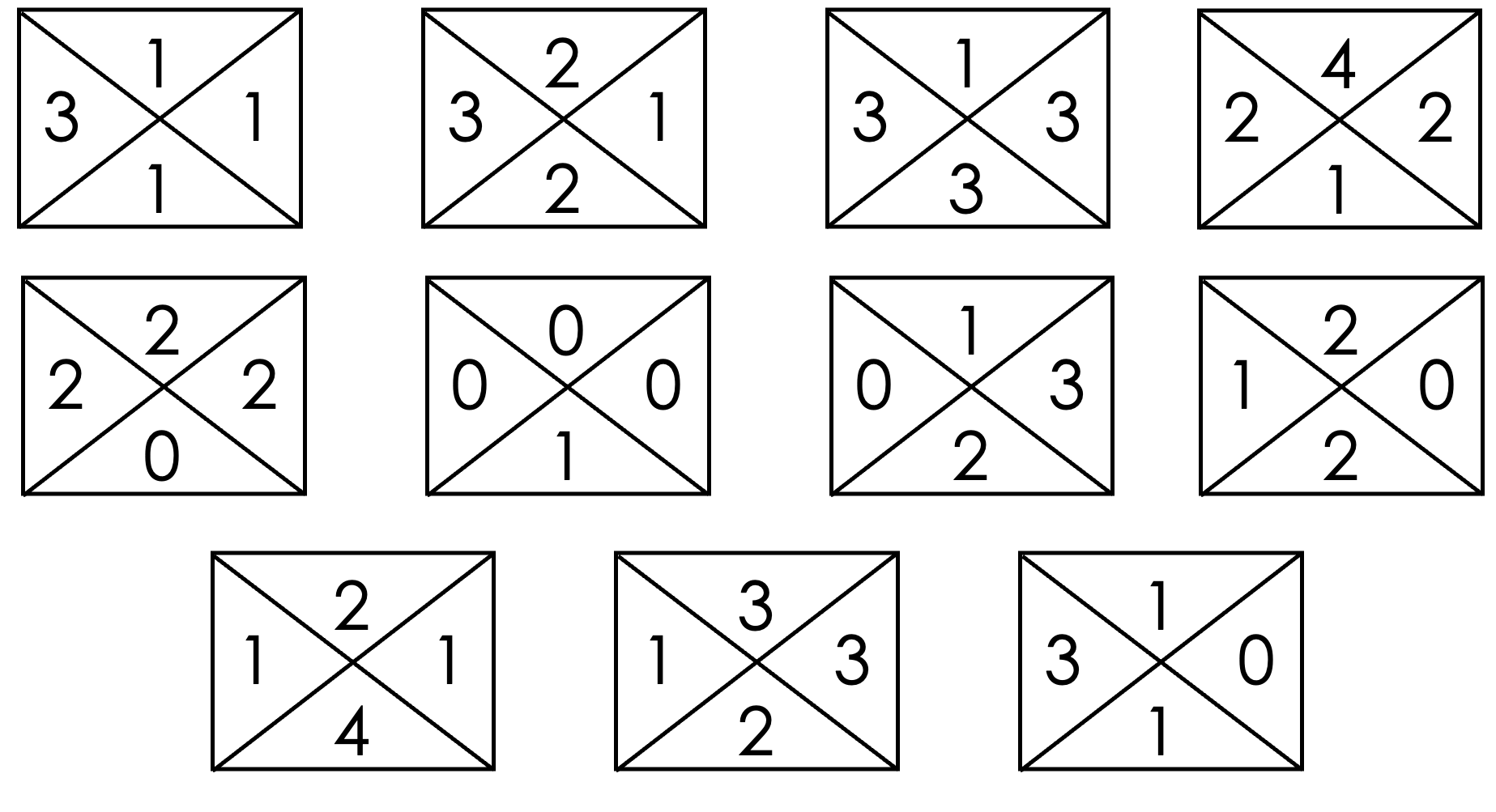}
		\caption{Alphabet G3}
		\label{fig:G3_analysis_a}
	\end{subfigure}
	\hfill 
	\begin{subfigure}[b]{0.32\linewidth}
		\centering
		\includegraphics[width=\linewidth]{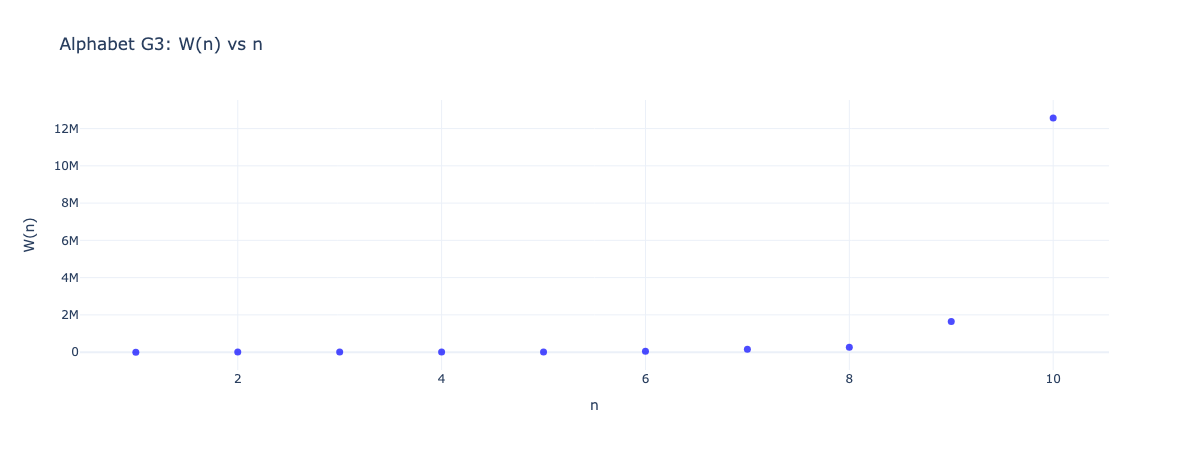}
		\caption{$W_{\Gamma}(n)$ vs $n$}
		\label{fig:G3_analysis_b}
	\end{subfigure}
	\hfill 
	\begin{subfigure}[b]{0.32\linewidth}
		\centering
		\includegraphics[width=\linewidth]{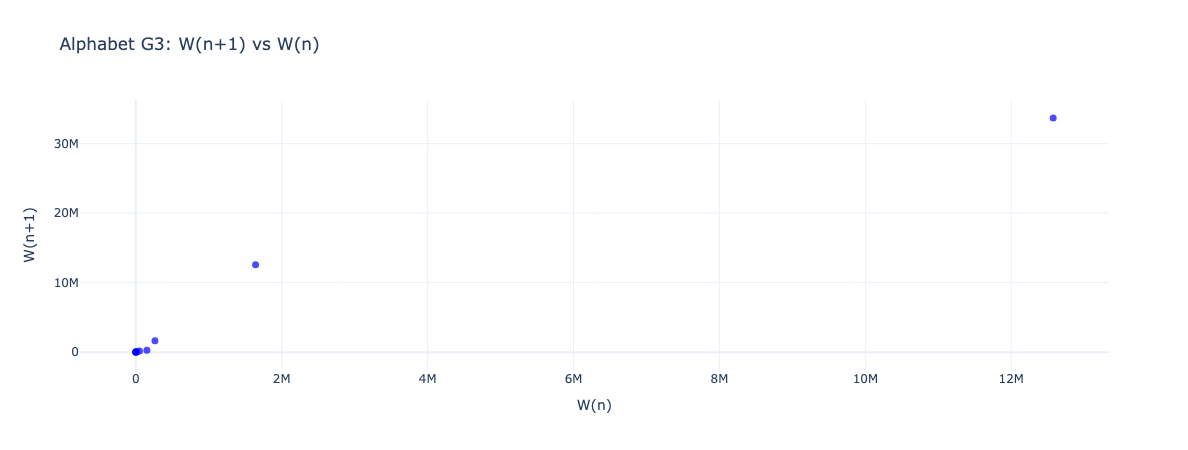}
		\caption{$W_{\Gamma}(n+1)$ vs $W_{\Gamma}(n)$}
		\label{fig:G3_analysis_c}
	\end{subfigure}
	
	\caption{Alphabet G3 and its associated complexity plots.}
	\label{fig:G3_analysis}
\end{figure}

\textbf{Alphabet G3.} Documented in \cite{[4]}, this set illustrates an
aperiodic system that, unlike the previous case, is devoid of self-similarity.
This structural difference introduces significant fluctuations in the growth
profile of $W_{\Gamma}(n)$, distinguishing it from the smooth hierarchy of G2.
Consequently, the phase-space diagram $W_{\Gamma}(n+1)$ versus $W_{\Gamma}(n)$
lacks a clear fractal signature. Despite these irregularities, the system
successfully ensures complete plane coverage without periodicity, justifying
its inclusion in the \textquotedblleft good\textquotedblright\ category while
offering a more intricate dynamical complexity. The visual analysis is shown
in Fig.~\ref{fig:G3_analysis}.

\begin{figure}[htbp]
	\centering
	\begin{subfigure}[b]{0.2\linewidth}
		\centering
		\includegraphics[width=\linewidth]{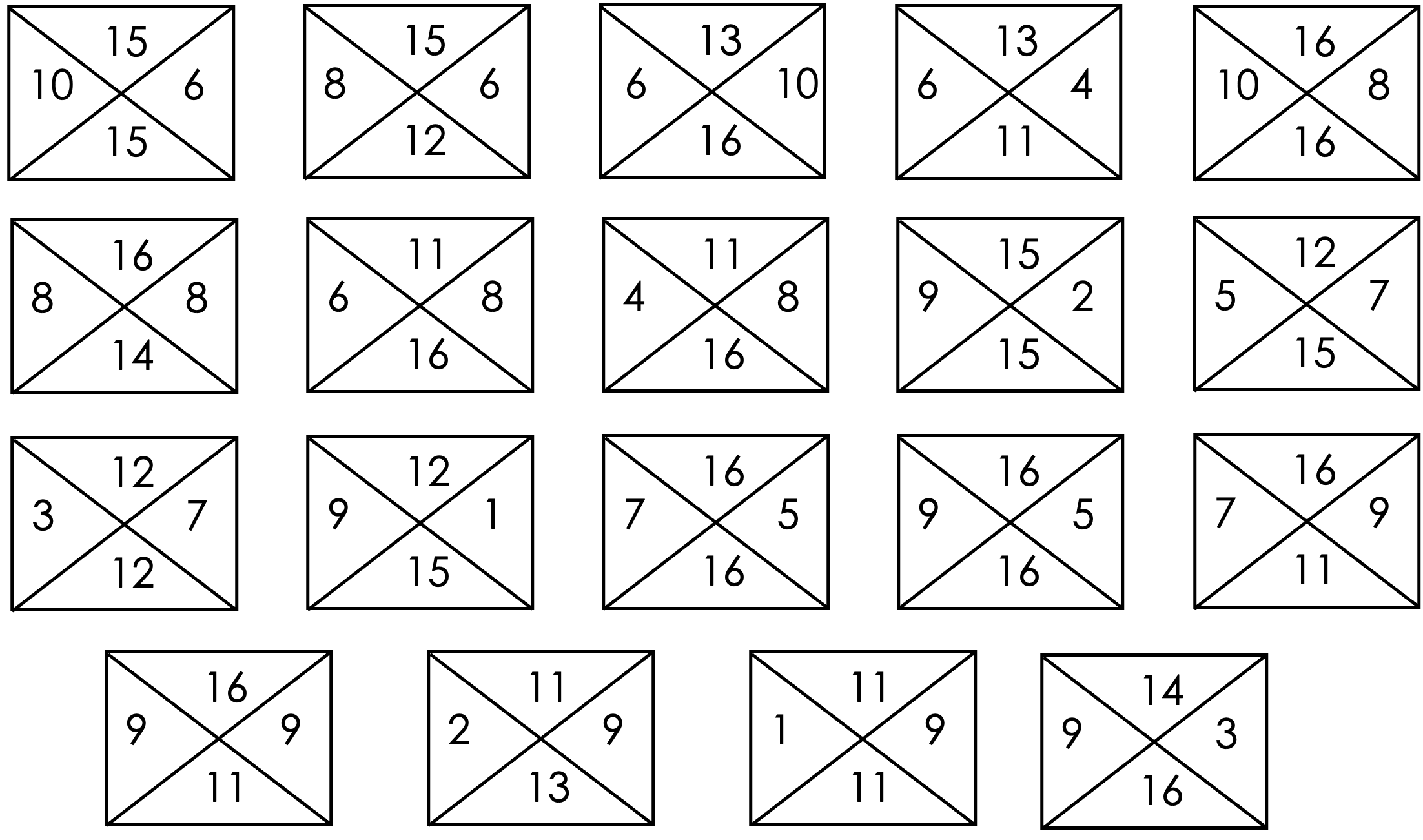}
		\caption{Alphabet G4}
		\label{fig:G4_analysis_a}
	\end{subfigure}
	\hfill 
	\begin{subfigure}[b]{0.32\linewidth}
		\centering
		\includegraphics[width=\linewidth]{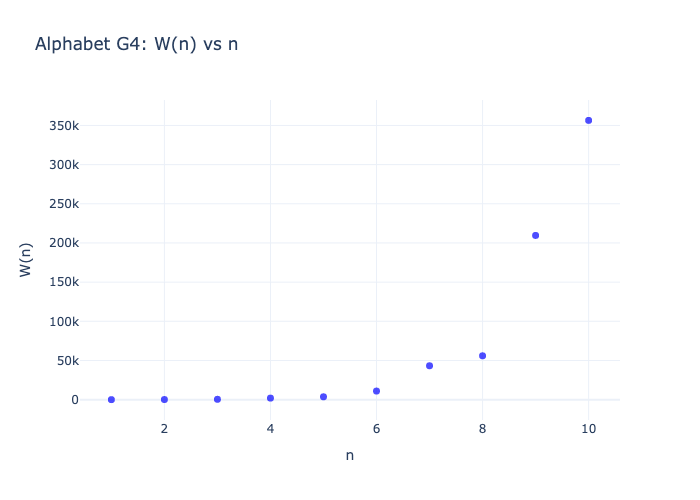}
		\caption{$W_{\Gamma}(n)$ vs $n$}
		\label{fig:G4_analysis_b}
	\end{subfigure}
	\hfill 
	\begin{subfigure}[b]{0.32\linewidth}
		\centering
		\includegraphics[width=\linewidth]{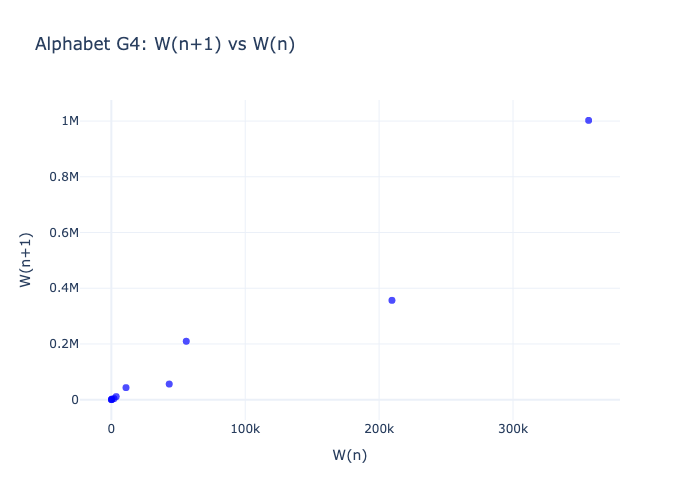}
		\caption{$W_{\Gamma}(n+1)$ vs $W_{\Gamma}(n)$}
		\label{fig:G4_analysis_c}
	\end{subfigure}
	
	\caption{Alphabet G4 and its associated complexity plots.}
	\label{fig:G4_analysis}
\end{figure}

\textbf{Alphabet G4.} As investigated in \cite{[4.1]}, this set introduces a
distinct dynamical regime characterized by high combinatorial entropy. The
complexity curve $W_{\Gamma}(n)$ exhibits a pronounced convex growth profile,
indicating a rapid proliferation of valid configurations as the system size
expands. This explosive behavior is mirrored in the phase-space projection
$W_{\Gamma}(n+1)$ versus $W_{\Gamma}(n)$, which traces a steep trajectory
distinct from the linear or fractal patterns observed in previous cases.
Despite this high density of states, the alphabet retains the structural
consistency required to classify it as \textquotedblleft
good\textquotedblright, offering a robust tiling capability with extensive
configurational freedom. The graphical breakdown is depicted in
Fig.~\ref{fig:G4_analysis}.

\begin{figure}[htbp]
	\centering
	\begin{subfigure}[b]{0.2\linewidth}
		\centering
		\includegraphics[width=\linewidth]{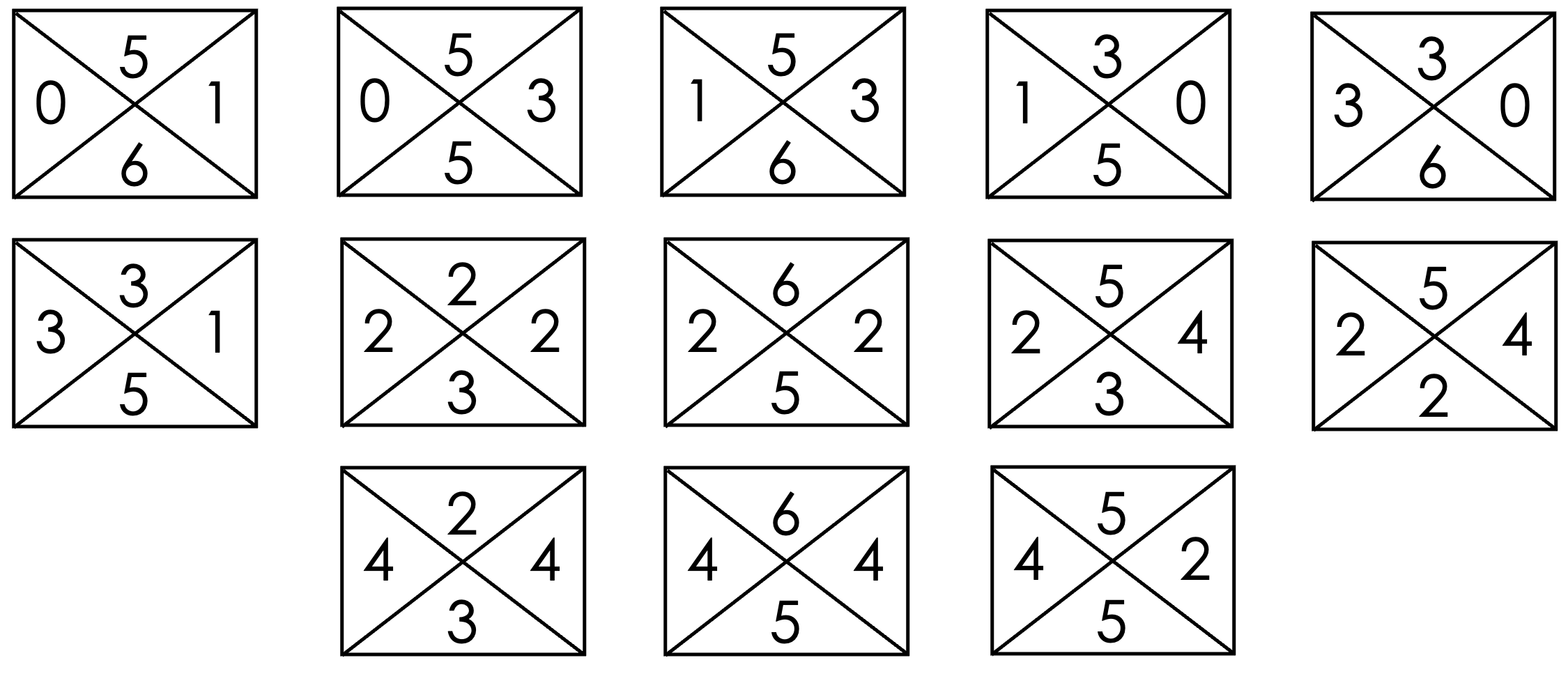}
		\caption{Alphabet G5}
		\label{fig:G5_analysis_a}
	\end{subfigure}
	\hfill 
	\begin{subfigure}[b]{0.32\linewidth}
		\centering
		\includegraphics[width=\linewidth]{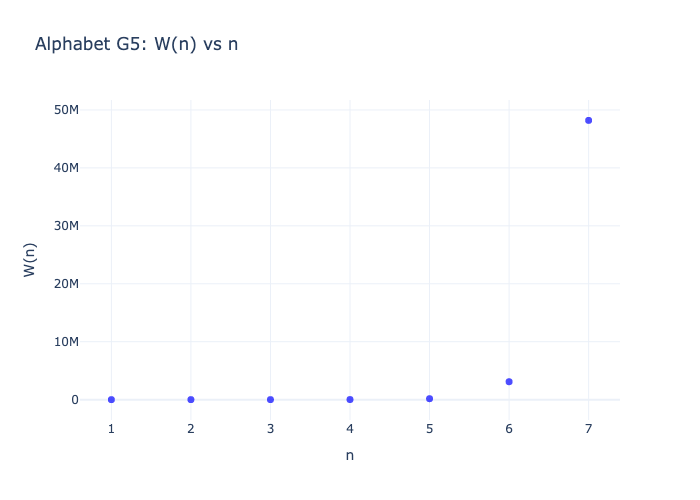}
		\caption{$W_{\Gamma}(n)$ vs $n$}
		\label{fig:G5_analysis_b}
	\end{subfigure}
	\hfill 
	\begin{subfigure}[b]{0.32\linewidth}
		\centering
		\includegraphics[width=\linewidth]{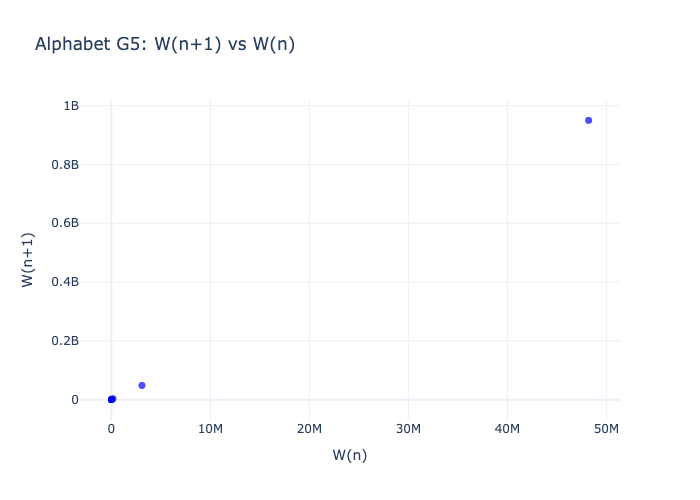}
		\caption{$W_{\Gamma}(n+1)$ vs $W_{\Gamma}(n)$}
		\label{fig:G5_analysis_c}
	\end{subfigure}
	
	\caption{Alphabet G5 and its associated complexity plots.}
	\label{fig:G5_analysis}
\end{figure}

\textbf{Alphabet G5.} As detailed in \cite{[23]}, this family represents an
extreme case of combinatorial richness within the \textquotedblleft
good\textquotedblright\ spectrum. The complexity metrics reveal a
hyper-expansive behavior: the $W_{\Gamma}(n)$ plot demonstrates a dramatic
surge, reaching magnitudes of tens of millions of configurations ($\sim5
\times10^{7}$) at very small system sizes ($n=7$). This explosive scaling is
further evidenced in the phase-space projection $W_{\Gamma}(n+1)$ versus
$W_{\Gamma}(n)$, where the trajectory ascends vertically towards the
billion-scale regime. Such behavior indicates a system with maximal entropic
density that, remarkably, preserves the tiling consistencies required for
validity. The graphical characterization is displayed in
Fig.~\ref{fig:G5_analysis}.

As suggested by the physical intuition, it is convenient to use the entropy
$S_{\Gamma}(n)$ as main variable for the plots:%

\begin{equation}
S_{\Gamma}(n)=\log_{10}(W(n)) \label{log}%
\end{equation}

The presence of the logarithm fixes the scale issue of $W_{\Gamma}(n)$ (see
\cite{MarcoFabrizio}).

Thus, the \textit{heuristic} proposed in \cite{MarcoFabrizio} is divided into
the following steps:

\textbf{Step 1)} Compute numerically as many values as possible of the
function $W_{\Gamma}\left(  n\right)  $ (subject to the available resources,
the ideal situation being having $n_{\max}$ larger than $q$ of, at least, a
factor of 2 or 3: see the comments below Eq. (\ref{protocol1a})) for the
alphabet $\Gamma$ of interest.

\textbf{Step 2)} Due to the large amount of combinations, it is better to use
the entropy $S_{\Gamma}\left(  n\right)  $ as main variable .

\textbf{Step 3)} Produce the plot $P=(X,Y)=\left(  S_{\Gamma}\left(  n\right)
,S_{\Gamma}\left(  n+1\right)  \right)  $ as described in the previous section.

\textbf{Step 4)} Construct the best fit of $S_{\Gamma}\left(  n+1\right)
=f_{\Gamma}\left(  S_{\Gamma}\left(  n\right)  \right)  $ and compute the
employ Kendall's Tau ($\tau$).

To validate the scalability of the system, we strictly adhere to the
methodological framework established in \cite{MarcoFabrizio}. This previous
work identified Kendall's Tau ($\tau$) as the most effective metric for
distinguishing between \textquotedblleft good" (predictable) and
\textquotedblleft bad" (unstable) alphabets in the context of Wang tiling
entropy. $\tau$ is defined as:%

\[
\tau= \frac{(\text{Number of concordant pairs}) - (\text{Number of discordant
pairs})}{\frac{1}{2} n (n - 1)}
\]

As demonstrated in \cite{MarcoFabrizio}, $\tau$ is robust against outliers and
excels at detecting monotonic trends without assuming strict linearity.
Consequently, we adopt their criterion: only alphabets exhibiting high $\tau$
values (indicating a regular, monotonic trajectory) are considered suitable
for asymptotic analysis.

\begin{figure}[ptbh]
\centering
\begin{subfigure}[b]{0.3\linewidth}
		\centering
		\includegraphics[width=\linewidth]{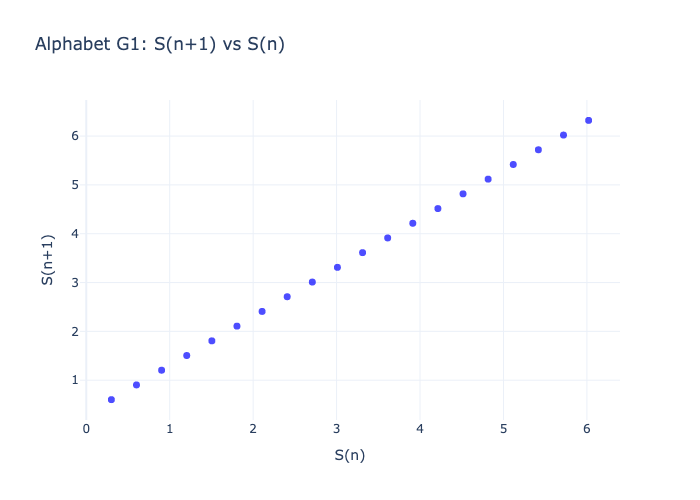}
		\caption{Family G1}
		\label{fig:reg_G1}
	\end{subfigure}
\hfill\begin{subfigure}[b]{0.3\linewidth}
		\centering
		\includegraphics[width=\linewidth]{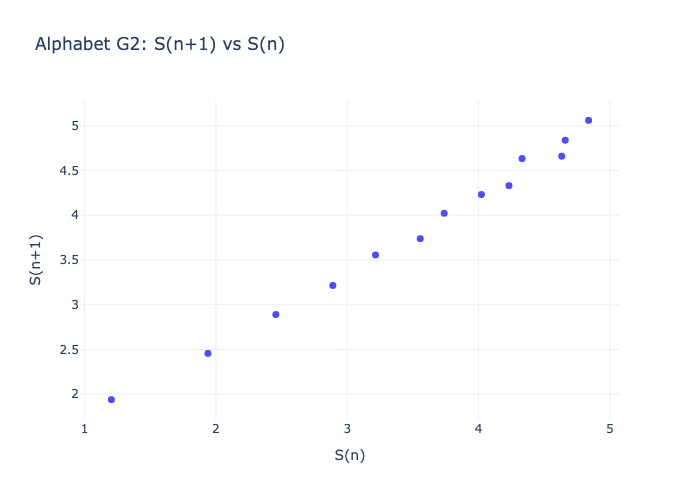}
		\caption{Family G2}
		\label{fig:reg_G2}
	\end{subfigure}
\hfill\begin{subfigure}[b]{0.3\linewidth}
		\centering
		\includegraphics[width=\linewidth]{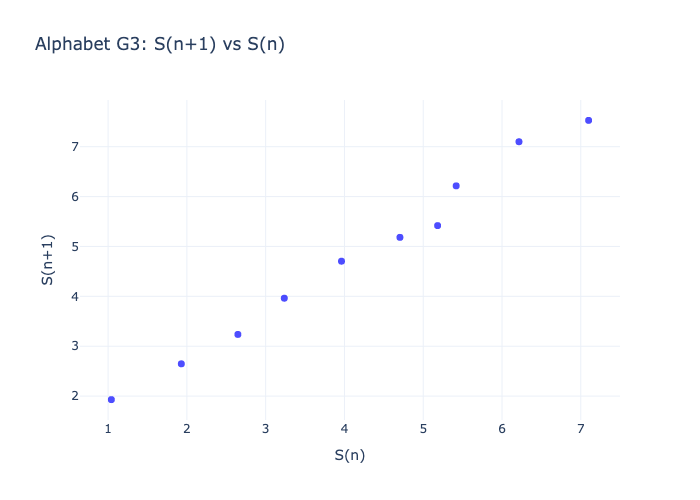}
		\caption{Family G3}
		\label{fig:reg_G3}
	\end{subfigure}
\par
\vspace{1em}
\par

\begin{subfigure}[b]{0.3\linewidth}
		\centering
		\includegraphics[width=\linewidth]{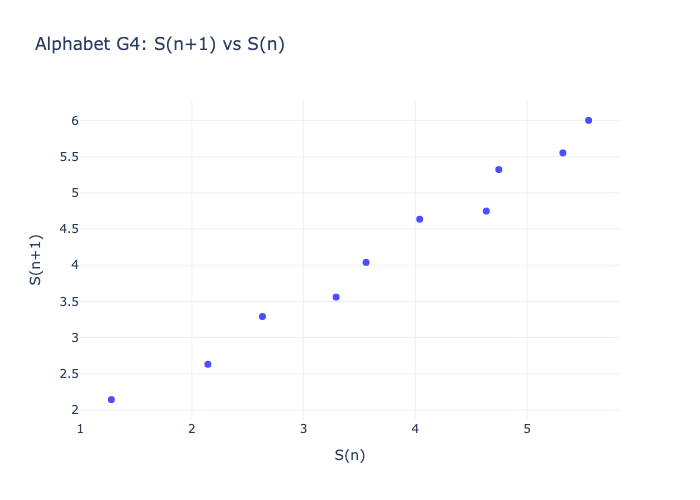}
		\caption{Family G4}
		\label{fig:reg_G4}
	\end{subfigure}
\hspace{0.05\linewidth}
\begin{subfigure}[b]{0.3\linewidth}
		\centering
		\includegraphics[width=\linewidth]{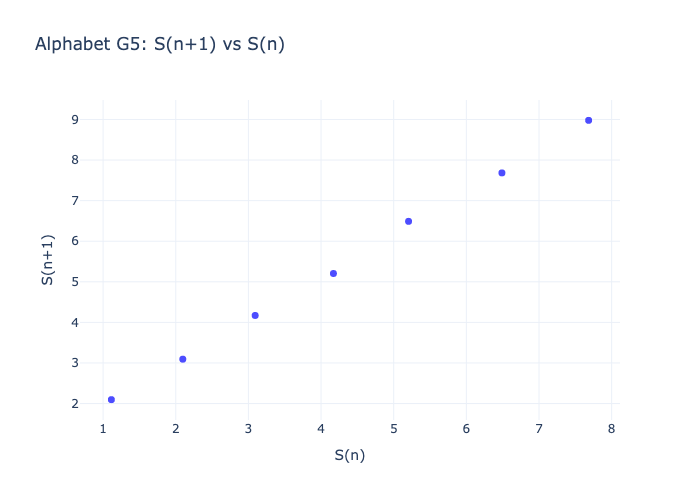}
		\caption{Family G5}
		\label{fig:reg_G5}
	\end{subfigure}
\caption{Regression dynamics ($S_{\Gamma}(n+1)$ vs $S_{\Gamma}(n)$) for the
analyzed alphabets. The plots illustrate the monotonic stability
characteristic of \textquotedblleft good" alphabets.}%
\label{fig:regression_G1_G5}%
\end{figure}

The empirical validation of the selected alphabets is substantiated by the
results shown in Fig. \ref{fig:regression_G1_G5} and Table \ref{tab:tabla_reg}%
. The figure illustrates the regression dynamics for alphabets G1 through G5,
demonstrating that all five adhere to the regular monotonic trajectory
characteristic of \textquotedblleft good" alphabets. This visual evidence is
quantitatively corroborated in Table \ref{tab:tabla_reg}, where the
consistently high $\tau$ values confirm the robustness and stability of our selection.

\begin{table}[ptbh]
\centering
\par
\begin{tabular}
[c]{cccc}\hline
Alphabet & $c_{0} $ & $\gamma$ & $\tau$\\\hline
G1 & 1.25 & 0.899 & 1.00\\
G2 & 1.49 & 0.754 & 1.00\\
G3 & 1.54 & 0.810 & 1.00\\
G4 & 1.52 & 0.782 & 1.00\\
G5 & 1.69 & 0.811 & 1.00\\\hline
\end{tabular}
\caption{Alphabets Regression coeff}%
\label{tab:tabla_reg}%
\end{table}

\section{Relations with the theory of discrete Chaos}

The second protocol highlights the deep connections between deterministic
chaos, NP completeness and undecidability (see \cite{15a0} \cite{15a}
\cite{15d} \cite{15e} and references therein). In particular, the mapping
\begin{equation}
S_{\Gamma}\left(  n+1\right)  =f_{\Gamma}\left(  S_{\Gamma}\left(  n\right)
\right)  , \label{dynsyst1}%
\end{equation}
can be analyzed using the theory of discrete dynamical system (the results in
the present sections do not depend on the detailed form of $f_{\Gamma}\left(
z\right)  $).

In particular, if $f_{\Gamma}\left(  z\right)  $ \textit{is monotone
non-decreasing,} the solutions of the above dynamical system will tend to the
solutions of the simpler (non-chaotic) dynamical system here below:
\begin{equation}
S_{\Gamma}\left(  n+1\right)  =f_{\Gamma}^{(0)}\left(  S_{\Gamma}\left(
n\right)  \right)  \ ,\ \ \ f_{\Gamma}^{(0)}\left(  z\right)  =c_{0}z^{\gamma
}\ , \label{dynsyst2}%
\end{equation}
according to which $W_{\Gamma}\left(  n\right)  $ grows exponentially with $n$
with subexponential corrections.

On the other hand, when $f_{\Gamma}\left(  z\right)  $ \textit{is not a
monotone function the dynamical system} in Eq. (\ref{dynsyst1}) can enter into
a chaotic phase. Very roughly, the more peaked is the local maximum, the
closer is the shape of $f_{\Gamma}\left(  z\right)  $ to a logistic map in the
chaotic phase. Thus, although the explicit analytic form of $f_{\Gamma}\left(
z\right)  $ is not available, the present results show that the maps
$f_{\Gamma}\left(  z\right)  $ associated to good alphabets do not manifest
chaotic tendency while the ones associated to bad alphabets manifest a clear
chaotic tendency.

These considerations also suggest \textit{the presence of a third category of
alphabets}, leading to the identification of \textit{the edge of chaos
region}. The alphabets belonging to this region do not satisfy our criteria to
be good, however their $f_{\Gamma}\left(  z\right)  $ are not irregular enough
to ensure that the corresponding discrete differential equations are actually
chaotic. These alphabets could have, for instance, oscillating behaviors in
their degeneracy $W_{\Gamma}(n)$ but on top of a monotonic tendency: from the
dynamical system analogy viewpoint, this region could correspond to the
bifurcations cascade before the chaotic region. Thus, these alphabets could be
examples of alphabets which tile the whole plane but do not satisfy our
criteria. Unfortunately, we have been unable to find examples of alphabets
belonging to this category.

At last, a very marked departure from monotonic growth is observed in the
\"{}%
bad alphabets%
\"{}
region. Even when initial iterations may exhibit an increasing behaviors of
$S_{\Gamma}(n)$ and of $W_{\Gamma}(n)$, after a certain number of steps a
clear inversion appear. This can be described by a a downward-opening
parabola. As far as the goals of the present manuscript is concerned, a form
of the type
\begin{equation}
S_{\Gamma}(n+1)\approx S_{0}-a[S_{\Gamma}(n)-b]^{2}\ ,
\end{equation}
where $S_{0}$, $a>0$, and $b$ are constants, captures the essence of this
behavior. This tendency towards a decreasing of $S_{\Gamma}(n+1)$ as function
of $S_{\Gamma}(n)$ appears to be a distinguishing feature of alphabets that
are unable to tile the plane: the similarity with the logistic map is clear.

\section{Theoretical Implications: The Structure of the Tiling Space}

\begin{figure}[th]
\centering
\includegraphics[width=0.6 \linewidth]{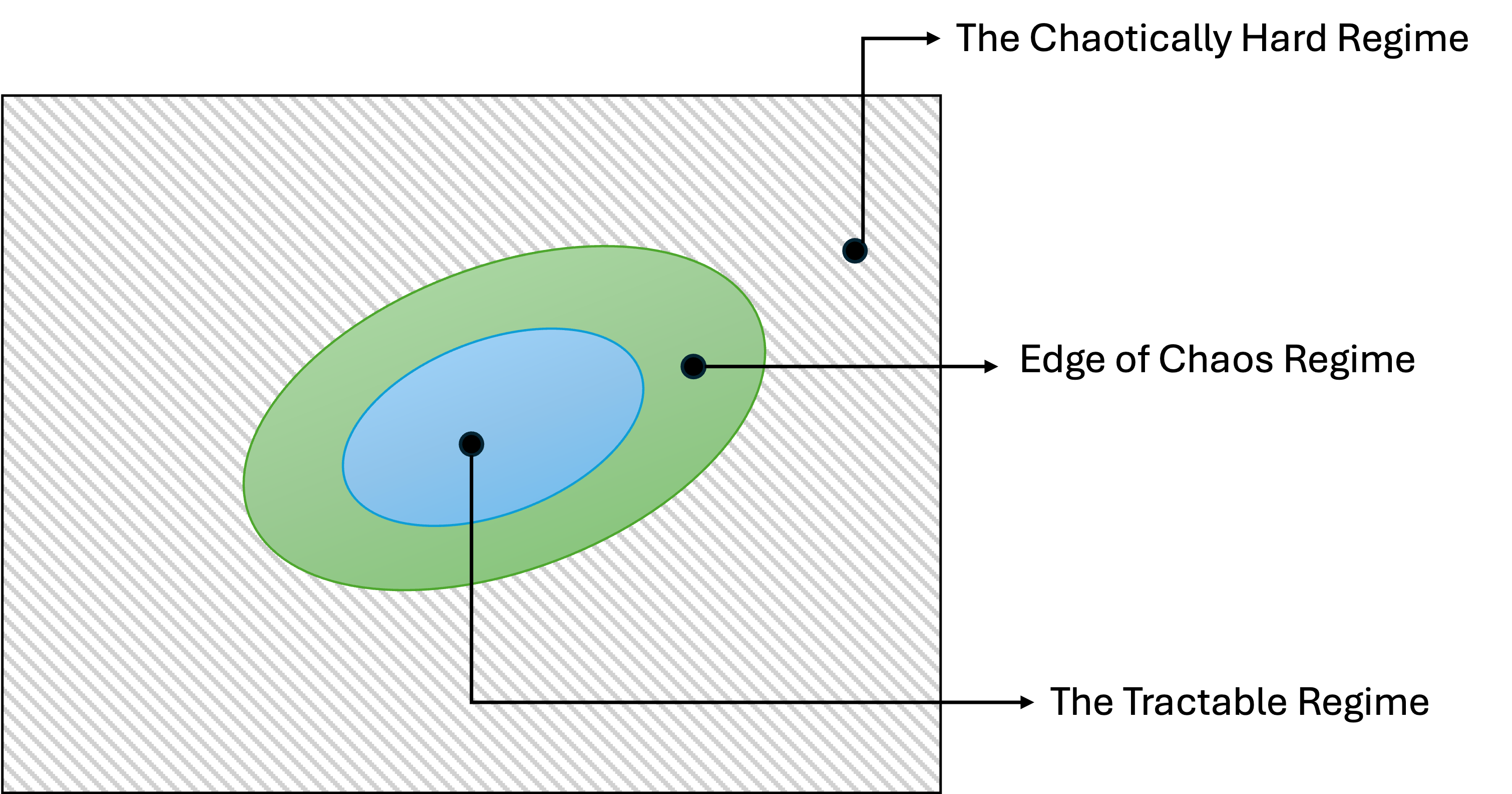}
\caption{Schematic representation of the configuration space bifurcation. The
diagram illustrates the division between the Tractable Regime (where
polynomial algorithms exist) and the Chaotically Hard Regime, separated by the
hypothetical intermediate Edge of Chaos Regime.}%
\label{regimenes}%
\end{figure}

The results presented in this work provide strong empirical support for the
structural bifurcation of the configuration space postulated in the
Introduction. Our successful implementation of the heuristic for specific
families confirms that the solution space is indeed non-homogeneous and can be
partitioned into the distinct regimes visualized in Fig. \ref{regimenes}:

\begin{enumerate}

\item \textbf{The Tractable Regime:} The performance of families G1 through G5
(Fig. \ref{fig:regression_G1_G5}) validates the existence of this sub-region.
Here, the topological structure allows for the construction of polynomial-time
algorithms, as demonstrated by our Chess/Go-inspired strategy. The high $\tau$
values observed indicate that asymptotic behavior in this regime is
predictable, stable, and computationally accessible.

\item \textbf{The Chaotically Hard Regime:} Conversely, the failure of the
heuristic strategy on unstable alphabets (those with low $\tau$ values) points
to the existence of this second region. In this regime, the \textquotedblleft
dressing\textquotedblright\ procedure fails because the local solutions at
step $n$ do not reliably predict the constraints at step $n+1$. The absence of
efficient algorithms here supports the hypothesis of computational
irreducibility, where the system's behavior is analogous to the chaotic phase
of a discrete dynamical system.

\item \textbf{The Edge of Chaos Regime:} While our current dataset focuses on
distinguishing the tractable from the hard, we maintain the hypothesis of a
third boundary region. Alphabets in this regime would satisfy neither the
stability criteria of the Tractable Regime nor the total unpredictability of
the Chaotic Regime. Identifying an explicit instance of an alphabet on this
\textquotedblleft edge\textquotedblright\ remains an open challenge for future research.
\end{enumerate}

\section{Polynomial Time Algorithms for Good Alphabets}

In this section, we point at the construction of a polynomial-time algorithm
specifically designed for \textquotedblleft good\textquotedblright\ alphabets.
The conceptual foundation of this strategy draws inspiration from the
strategic dynamics found in games such as chess or Go. In these domains, a
decisive advantage, or victory is often secured through a calculated sacrifice
of material (e.g., trading a rook for a positional advantage or placing a
stone in hostile territory) to yield a substantial long-term benefit.

Analogously, the present algorithmic strategy is predicated on a
\textit{strategic computational sacrifice}. We accept a significant loss of
efficiency during the initial stages to secure a \textit{massive computational
gain} in the asymptotic regime. This initial sacrifice is necessitated by two
fundamental requirements. First, it is imperative to determine whether the
alphabet $\Gamma$ qualifies as \textquotedblleft good\textquotedblright%
\ according to the heuristic criteria. Second, we must identify the specific
set of \textit{seed solutions} for small $r \times r$ squares (where $r
\approx n_{\max}$, the threshold required to establish the alphabet's
quality). These seeds constitute the boundary conditions necessary to
construct the tiling of the macroscopic $N \times N$ square, where $N \gg
n_{\max}$.

The efficacy of this strategy relies on a key empirical observation: good
alphabets exhibit a distinct regularity in the transition from $n \times n$
configurations to $(n+1) \times(n+1)$ configurations. From a practical
standpoint, this regularity implies that \textit{the solution space at the
$(n+1)$-th step can be effectively reconstructed using the solutions from the
$n$-th step}. This observation is the mechanism responsible for the
aforementioned long-term benefit. Indeed, if the validity of a configuration
at step $(n+1)$ is conditional on the configuration at step $n$, the
algorithmic complexity shifts from an exponential search to a polynomial
trajectory. Specifically, our data suggests that from this transition point
onward, the cost scales linearly. Consequently, the proposed strategy entails
a hybrid cost structure: a fixed initial cost followed by a polynomial growth.
Crucially, the initial cost depends solely on the cardinality $q = |\Gamma|$
and can indeed be substantial, but is \textit{independent of the target system
size} $N$.

\subsection{The Algorithmic Procedure}

Based on the principles outlined above, the strategy to explicitly construct a
Wang tiling for an $N \times N$ square is structured into two main phases
comprising the following sequential steps:

\subsubsection*{Phase I: Primitive Strategy}

\textbf{Step 1: Identification and Seed Generation.} The primary objective of
this step is to certify $\Gamma$ as a \textquotedblleft good\textquotedblright%
\ alphabet. This process requires $n_{\max}$ iterations, where $n_{\max}\gg
q$. It is vital to emphasize that $n_{\max}$ is an intrinsic parameter of the
alphabet and does not depend on the target size $N$. During this phase, we
explicitly compute $W_{\Gamma}(n_{\max})$ tilings for $n_{\max}\times n_{\max
}$ squares. Let us denote this quantity as $W=W_{\Gamma}(n_{\max})$. These $W$
configurations represent our \textit{seed solutions}. While the computational
cost of this step is high because it necessitates a brute-force approach
representing the \textquotedblleft sacrifice\textquotedblright\ of efficiency
it remains a fixed overhead, denoted as $C(\Gamma)$, which is constant
regardless of how large $N$ becomes. Note that, according to our numerical
results, the strategy often requires only a small subset of these $W$ seeds to
successfully propagate the tiling.

\subsubsection*{Phase II: Advanced Strategy}

\textbf{Step 2: Heuristic Expansion.} Once the $W$ seed solutions are
established, we proceed to construct the tilings for the $(n_{\max}+1)
\times(n_{\max}+1)$ square. By the definition of a good alphabet, the number
of valid solutions at step $(n+1)$ is strictly greater than or equal to the
number of solutions at step $n$. Therefore, it is reasonable to hypothesize
that the tilings at step $(n+1)$ can be generated by \textquotedblleft
dressing\textquotedblright\ the existing tilings from step $n$. In this
context, \textit{dressing} is defined as taking a valid tiling of an $n_{\max}
\times n_{\max}$ square and completing its perimeter to form a valid
$(n_{\max}+1) \times(n_{\max}+1)$ configuration.

This concept constitutes the core of our strategy. Once we have established
with sufficient confidence\footnote{As previously emphasized, absolute
certainty regarding an alphabet's classification is theoretically elusive.
However, our proposed heuristic has proven to be extremely effective in
practice.} that the alphabet is good and that the system has entered a
monotonically increasing phase of $W_{\Gamma}(n)$, we abandon the search for
$(n+1)$ solutions from scratch. Instead, we build directly upon the available
$n \times n$ topologies. This procedure yields a massive reduction in
computational time. Our experimental data confirms that the cost of Step 2 is
polynomial in time, as the number of combinatorial possibilities to check is
drastically reduced by the boundary constraints of the seed solutions.

\textbf{Step 3: Iterative Propagation.} The dressing procedure described in
Step 2 is iterated for all $n > n_{\max}$ until the final target size $N$ is reached.

\vspace{1em}

\begin{figure}[th]
\centering
\includegraphics[width=\linewidth]{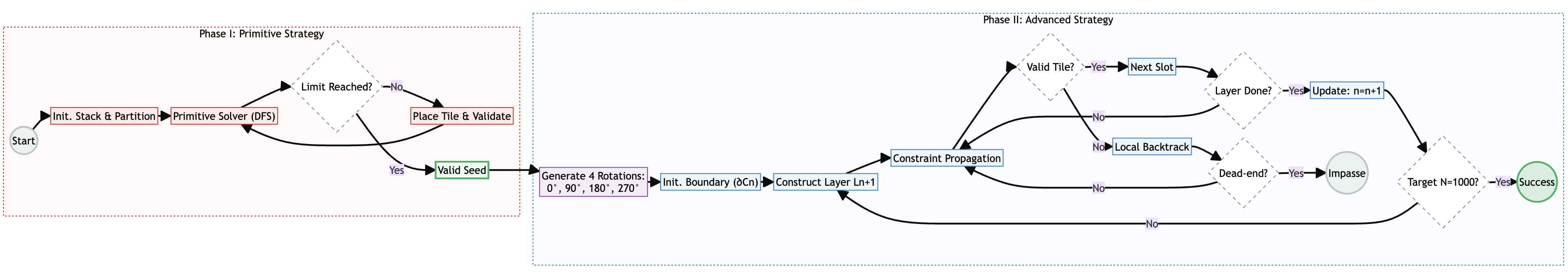}  \caption{Schematic
representation of the Hybrid Algorithm, illustrating the transition from the
Primitive Strategy (DFS) to the Advanced Strategy (Layered Growth).}%
\label{fig:Algorithm}%
\end{figure}

To provide a holistic view of the operational logic, the complete workflow is
schematized in Figure \ref{fig:Algorithm}. This diagram visually articulates
the sequential progression through the strategy, highlighting the structural
bifurcation between the exhaustive search (Primitive Strategy) used for seed
generation (Phase I) and the boundary-driven heuristic employed for asymptotic
mapping (Advanced Strategy, Phase II). The flowchart explicitly maps the
critical transition point where the algorithm shifts from combinatorial brute
force to linear-time layered growth, detailing the decision loops required to
validate the topological extensibility of the selected seeds up to the target scale.

\subsection{Hardware Limitations and Constraints}

While theoretically sound, the practical implementation of \textbf{Step 1}
encounters limitations imposed by available hardware resources. In our
specific case, utilizing standard computing equipment (two laptops), we face
two primary constraints. First, the raw computing power limits the depth of
the initial brute-force search. Second, and more critically, the storage
required for the $W$ seeds can easily exceed several terabytes, saturating the
available disk capacity.

These constraints imply that we cannot practically achieve the ideal condition
of $n_{\max} \gg q$; instead, we are limited to an operational range of
$n_{\max} \approx2q$. Theoretically, this poses a challenge, as a larger
$n_{\max}$ would provide a higher statistical guarantee that at least one $n
\times n$ tiling can be successfully dressed to $(n+1)$. Furthermore, the
magnitude of $W$ (the number of valid seeds) at $n_{\max} \approx2q$ is often
so large that we cannot computationally test every single seed, forcing us to
rely on a subset. This limitation likely explains why the algorithm works
explicitly only for a subset of good alphabets in our tests. Nevertheless,
despite these hardware-imposed bounds, the strategy has demonstrated
remarkable success across a range of good alphabets found in the literature.

\section{Computational Performance and Scalability Analysis}

In this section, we present the empirical validation of the Hybrid Algorithm
applied to the analyzed alphabets (G1--G5). To quantify the efficiency of the
proposed heuristic within the \textit{Tractable Regime}, we focus on two
critical performance metrics: the computational time required to advance the
tiling from size $n$ to $n+1$, and the peak memory consumption associated with
processing the boundary constraints.

The results presented below explicitly illustrate the operational phase
transition described in the methodology. Specifically, the data highlights the
sharp contrast between the potential growth inherent to the \textit{Primitive
Strategy} (Phase I) and the polynomial and empirically linear scaling achieved
by the \textit{Advanced Strategy} (Phase II). This quantitative analysis
confirms that for the selected "good" alphabets, the initial computational
sacrifice effectively secures access to an asymptotic regime where the tiling
process becomes computationally efficient and predictable.

\subsection{Performance Analysis: Alphabet G1}

\begin{figure}[ptbh]
\centering
\begin{subfigure}[c]{0.2\linewidth}
		\centering
		\includegraphics[width=\linewidth]{figures/Alphabet_G1.png}
		\caption{Alphabet G1}
		\label{fig:G1_a}
	\end{subfigure}
 \hfill
\hspace{2em}
\begin{subfigure}[c]{0.60\linewidth}
		\centering
		\includegraphics[width=\linewidth]{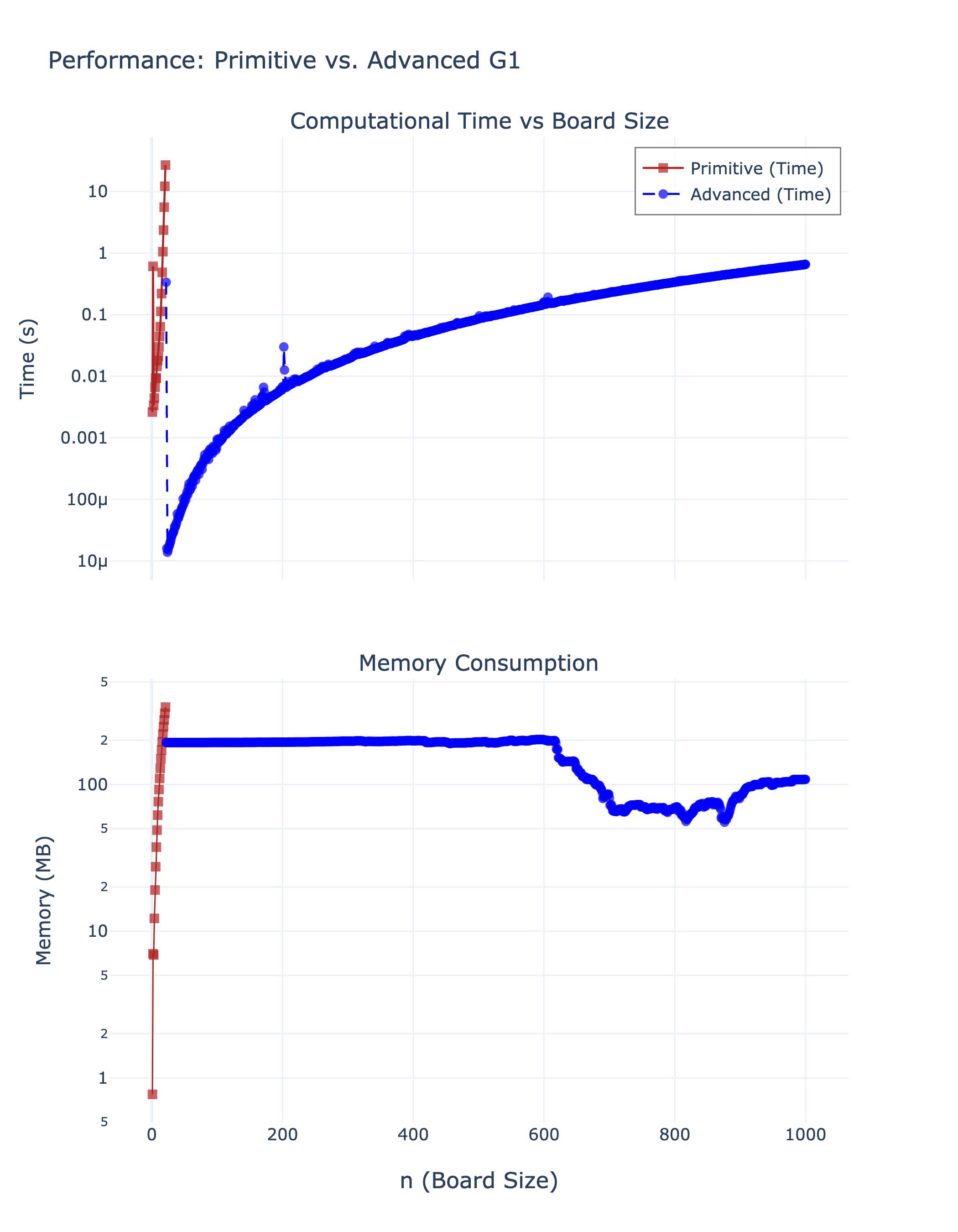}
		\caption{Time and Memory vs $n$}
		\label{fig:G1_b}
	\end{subfigure}
\caption{Performance Analysis: Alphabet G1.}%
\label{fig:G1_comp}%
\end{figure}

The computational behavior of the G1 family, as illustrated in
Fig.~\ref{fig:G1_comp}, provides a clear baseline for evaluating the proposed
hybrid strategy.

\textbf{Primitive Strategy (Exhaustive Search):} The red trajectory indicates
the performance of the brute-force enumeration. An initial spike in time and
memory consumption is observed at small $N$, attributable to the overhead of
data structure initialization and family loading rather than combinatorial
complexity. Following this setup phase, the cost grows exponentially. The
method hits a hard computational barrier at approximately $N \approx20$, where
both time and memory demands diverge vertically, rendering further growth via
this strategy unfeasible.

\textbf{Advanced Strategy (Layered Growth):} Upon switching to the layered
heuristic (blue trajectory), the system demonstrates remarkable scalability,
successfully reaching the target size of $N=1000$. The time complexity shifts
from exponential to a polynomial regime, maintaining execution times below 1
second even at maximal system size. Crucially, the memory profile remains
exceptionally stable, hovering around a constant baseline (approx. 2-5 MB) for
the majority of the process. This plateau suggests that for simple alphabets
like G1, the layered heuristic effectively optimizes the boundary storage,
requiring significant resources only during the initialization and
finalization stages, with minor fluctuations observed beyond $N=600$ likely
due to internal memory management or boundary readjustments.

\subsection{Performance Analysis: Alphabet G2}

\begin{figure}[ptbh]
\centering
\begin{subfigure}[c]{0.20\linewidth}
		\centering
		\includegraphics[width=\linewidth]{figures/Alphabet_G2.png}
		\caption{Alphabet G2}
		\label{fig:G2_a}
	\end{subfigure}

\hspace{2em}
\begin{subfigure}[c]{0.60\linewidth}
		\centering
		\includegraphics[width=\linewidth]{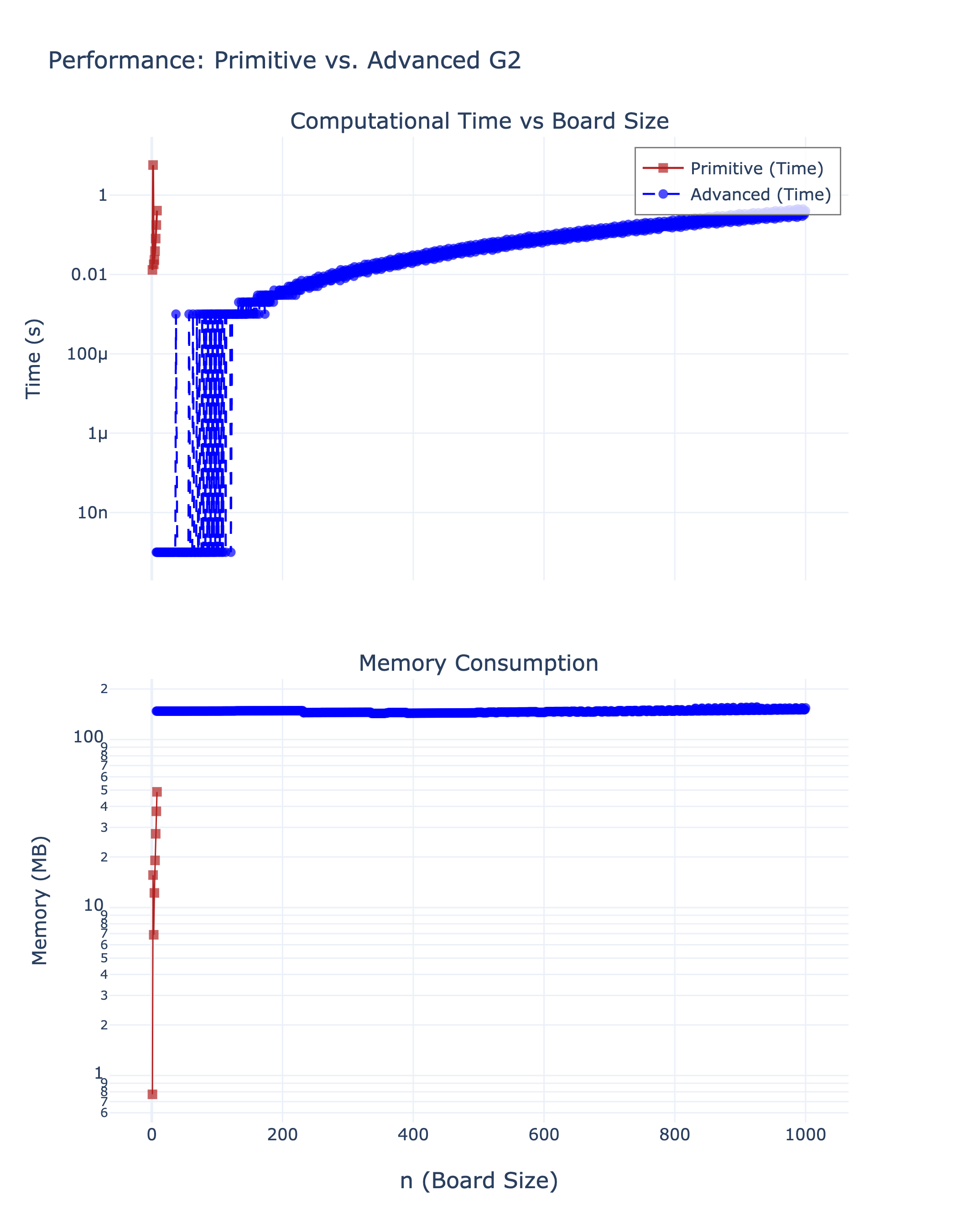}
		\caption{Time and Memory vs $n$}
		\label{fig:G2_b}
	\end{subfigure}
\caption{Performance Analysis: Alphabet G2.}%
\label{fig:G2_comp}%
\end{figure}

The computational dynamics of Alphabet G2, as detailed in
Fig.~\ref{fig:G2_comp}, underscore the critical advantage of the layered
heuristic when applied to complex, aperiodic systems ($|\Gamma|=14$).

\textbf{Primitive Strategy:} The limitations of the exhaustive search are
immediately exposed by the high cardinality of the tile set. The combinatorial
explosion occurs almost instantaneously; the red trajectory shows a vertical
divergence in computational cost, hitting the resource ceiling at a negligible
system size ($N < 5$). This confirms that for larger alphabets with complex
matching rules, the brute-force approach is computationally intractable from
the onset.

\textbf{Advanced Strategy:} Conversely, the layered growth strategy
successfully mitigates this complexity, achieving the target scale of $N=1000$.

\begin{itemize}
\item \textit{Time Efficiency:} The execution time scales polynomially,
remaining computationally viable (under 5 seconds) even at $N=1000$. The
visible fluctuations (noise) along the blue curve, more pronounced than in G1,
reflect the non-trivial local adjustments and backtracking required to satisfy
the aperiodic matching constraints of this specific family.

\item \textit{Memory Profile:} The memory behavior exhibits a distinct
\textquotedblleft static saturation\textquotedblright\ pattern. Due to the
large number of tiles, the system initializes with a high baseline memory
footprint ($\sim150$ MB). Remarkably, this consumption remains virtually
constant throughout the entire growth process. This plateau indicates that the
memory overhead is dominated by the initial storage of the transition rules
and the active boundary layer, rather than scaling with the total map area,
demonstrating excellent space complexity management for complex alphabets.
\end{itemize}

\subsection{Performance Analysis: Alphabet G3}

\begin{figure}[ptbh]
\centering
\begin{subfigure}[c]{0.20\linewidth}
		\centering
		\includegraphics[width=\linewidth]{figures/Alphabet_G3.png}
		\caption{Alphabet G3}
		\label{fig:G3_a}
	\end{subfigure}

\hspace{2em}
\begin{subfigure}[c]{0.60\linewidth}
		\centering
		\includegraphics[width=\linewidth]{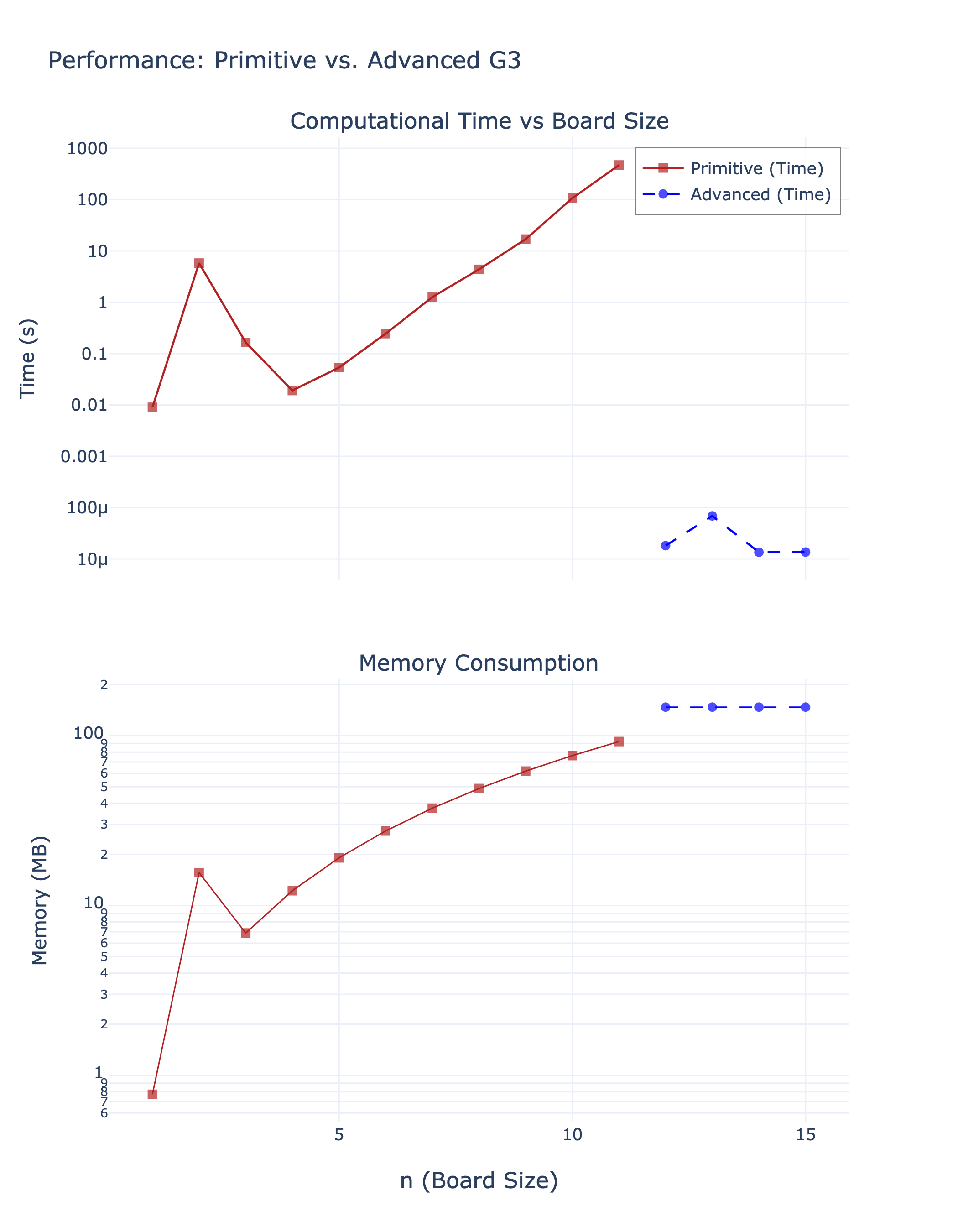}
		\caption{Time and Memory vs $n$}
		\label{fig:G3_b}
	\end{subfigure}
\caption{Performance Analysis: Alphabet G3.}%
\label{fig:G3_comp}%
\end{figure}

The analysis of Alphabet G3, presented in Fig.~\ref{fig:G3_comp}, reveals a
distinct computational boundary where the topological complexity of the tile
set challenges both search strategies. Unlike the previous cases, this family
exhibits a resistance to scalability that precludes mapping the system to
large $N$.

\textbf{Primitive Strategy:} The exhaustive enumeration (red trajectory)
follows a steep exponential curve typical of NP-hard tiling problems. After
the characteristic initialization spike at $N=2$, the computational time
degrades rapidly, exceeding 500 seconds merely to reach $N=11$. The memory
consumption mirrors this trend, growing linearly in the log-scale
(exponentially in real terms) until the process becomes intractable.

\textbf{Advanced Strategy:} In this specific instance, the transition to the
layered heuristic (blue trajectory) provides only a marginal extension of the
solvable domain.

\begin{itemize}
\item \textit{Early Saturation:} The heuristic attempts to pick up the mapping
from the limit of the primitive search ($N \approx12$) but encounters a severe
computational bottleneck almost immediately. The strategy manages to advance
the boundary only up to $N \approx15$ before stalling.

\item \textit{Algorithmic Stagnation:} This premature termination suggests
that the internal structure of G3?previously described as aperiodic but
non-self-similar?imposes rigid boundary constraints. These constraints likely
force the heuristic into local minima where no valid layer addition is
possible without extensive global backtracking, which the greedy nature of the
layered approach seeks to avoid.

\item \textit{Resource High-Water Mark:} The memory usage remains high and
static ($>100$ MB) for the few successful steps, indicating that maintaining
the complex boundary state of G3 is resource-intensive even for small system sizes.
\end{itemize}

\subsection{Performance Analysis: Alphabet G4}

\begin{figure}[ptbh]
\centering

\begin{subfigure}[c]{0.20\linewidth}
		\centering
		\includegraphics[width=\linewidth]{figures/Alphabet_G4.png}
		\caption{Alphabet G4}
		\label{fig:G4_a}
	\end{subfigure}

\hspace{2em}
\begin{subfigure}[c]{0.60\linewidth}
		\centering
		\includegraphics[width=\linewidth]{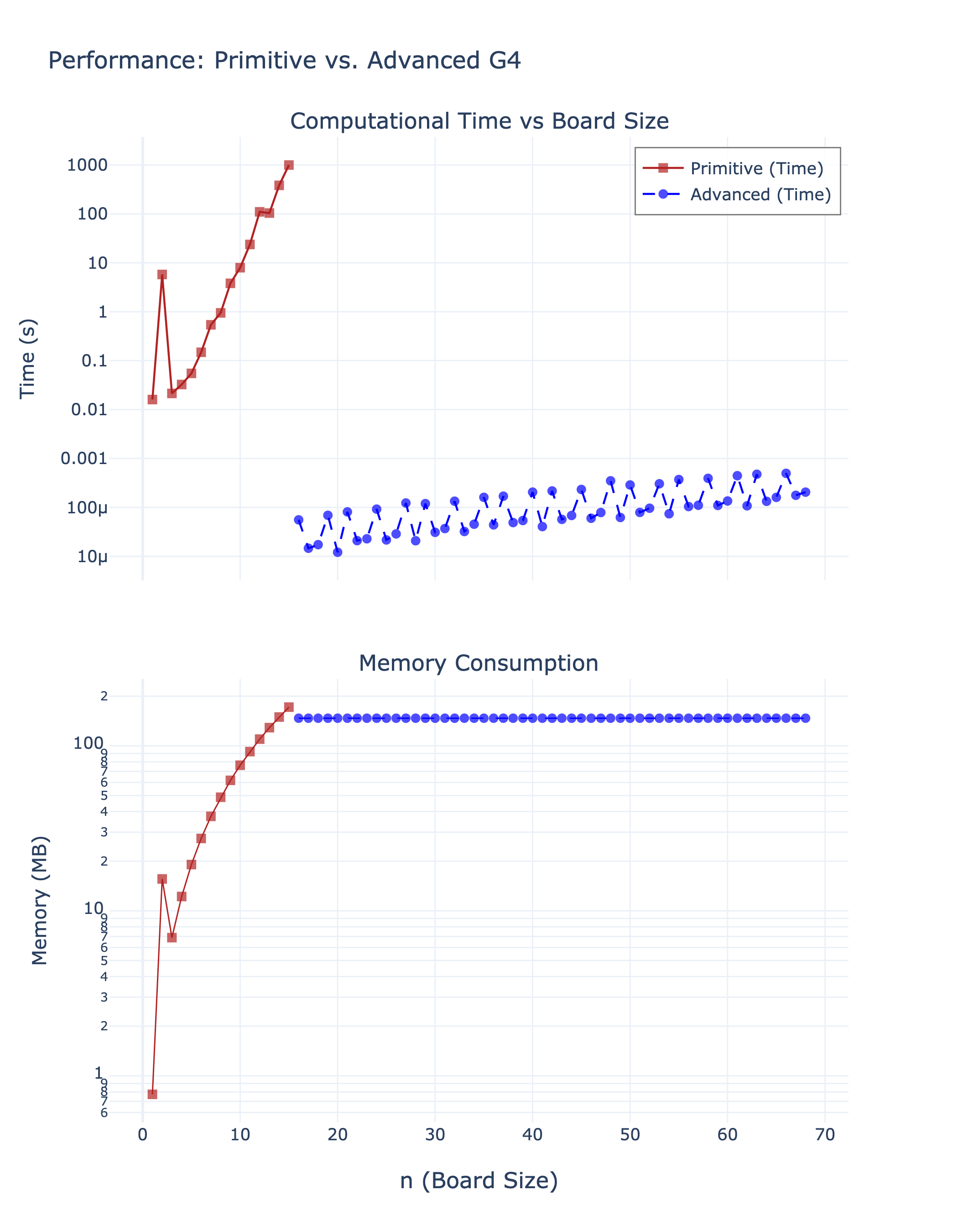}
		\caption{Time and Memory vs $n$}
		\label{fig:G4_b}
	\end{subfigure}
\caption{Performance Analysis: Alphabet G4.}%
\label{fig:G4_comp}%
\end{figure}

The computational trajectory of Alphabet G4, shown in Fig.~\ref{fig:G4_comp},
illustrates a scenario of extreme algorithmic acceleration bounded by a
delayed topological horizon.

\textbf{Primitive Strategy:} Given the large size of the alphabet
($|\Gamma|=16$), the brute-force approach faces a severe combinatorial wall.
The time complexity (red trajectory) scales vertically on the logarithmic
axis, exceeding $10^{3}$ seconds to resolve layer $n=14$. Similarly, memory
consumption rises exponentially, reaching saturation ($\approx150$ MB)
rapidly. This behavior confirms that for high-entropy alphabets, exact
enumeration is viable only for very small system sizes.

\textbf{Advanced Strategy:} The transition to the layered heuristic triggers a
dramatic phase shift in performance, extending the mapping capability up to
$n=68$.

\begin{itemize}
\item \textit{Hyper-Efficiency in Time:} Unlike the previous cases, the
execution time for the advanced strategy (blue points) drops orders of
magnitude, effectively hovering near the resolution floor of the timer ($t <
10^{-3}$ s). This indicates that, for the solvable range $n \in[15, 68]$, the
local matching rules are loose enough for the heuristic to find valid
configurations almost instantaneously without backtracking.

\item \textit{Memory Plateau:} The memory footprint exhibits a perfect
\textquotedblleft static plateau\textquotedblright. It inherits the high
baseline ($\sim150$ MB) established by the primitive phase but incurs zero
marginal cost for subsequent layers.

\item \textit{Topological Impasse:} Despite this efficiency, the growth
abruptly terminates at $n=68$. This suggests that while G4 allows for rapid
local expansion, it eventually accumulates global geometric constraints that
create an unsolvable boundary for a greedy approach. This represents a
\textquotedblleft delayed dead-end,\textquotedblright\ extending significantly
further than G3 but falling short of the asymptotic target ($N=1000$).
\end{itemize}

\subsection{Performance Analysis: Alphabet G5}

\begin{figure}[ptbh]
\centering
\begin{subfigure}[c]{0.20\linewidth}
		\centering
		\includegraphics[width=\linewidth]{figures/Alphabet_G5.png}
		\caption{Alphabet G5}
		\label{fig:G5_a}
	\end{subfigure}

\hspace{2em}

\begin{subfigure}[c]{0.60\linewidth}
		\centering
		\includegraphics[width=\linewidth]{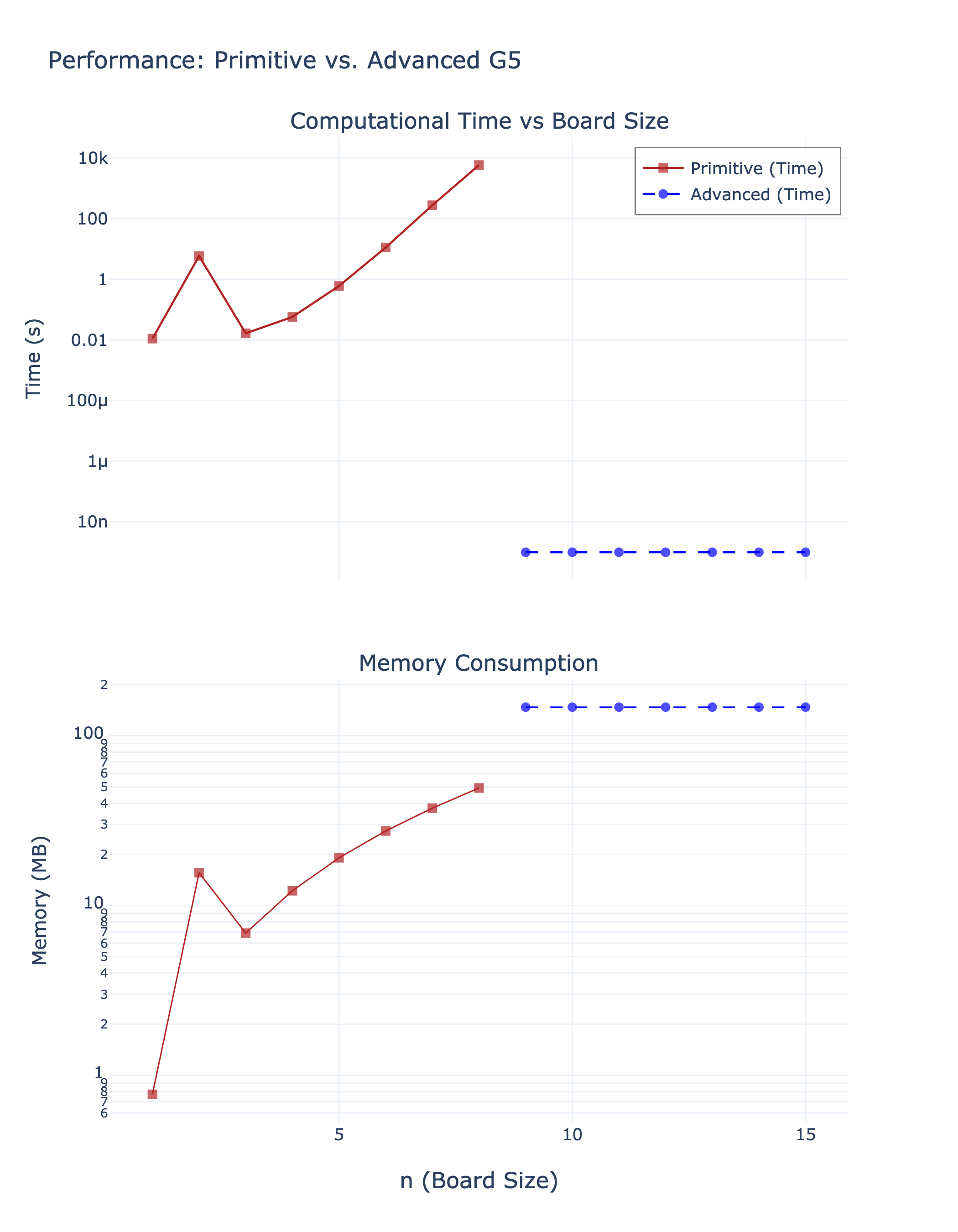}
		\caption{Time and Memory vs $n$}
		\label{fig:G5_b}
	\end{subfigure}
\caption{Performance Analysis: Alphabet G5.}%
\label{fig:G5_comp}%
\end{figure}

The computational assessment of Alphabet G5, depicted in
Fig.~\ref{fig:G5_comp}, represents the ultimate stress test for the proposed
algorithms, driven by the hyper-expansive entropic nature previously
identified in the complexity analysis.

\textbf{Primitive Strategy:} The brute-force enumeration faces a catastrophic
combinatorial divergence. Unlike previous cases where the time barrier was
encountered in the range of hundreds of seconds, G5 forces the primitive
algorithm into a prohibitive temporal regime, reaching $\sim10^{4}$ seconds
merely to resolve the system at $n=8$. The vertical scaling of both time and
memory confirms that the density of valid states for this alphabet renders
exact enumeration futile beyond trivial sizes.

\textbf{Advanced Strategy:} The implementation of the layered heuristic allows
the system to bypass the initial barrier, yet it encounters a secondary
\textquotedblleft entropic horizon\textquotedblright\ at $n=14$.

\begin{itemize}
\item \textit{Limited Extension:} Although the heuristic strategy manages to
initialize the boundary growth where the primitive method collapsed ($n
\approx9$), it only sustains the expansion for a short interval, terminating
at $n=14$. This suggests that the sheer volume of valid local
configurations?previously quantified in the tens of millions?overwhelms the
heuristic's decision-making process.

\item \textit{Resource Saturation:} The memory profile (blue trajectory)
remains static at a high saturation point ($\sim150$ MB). Unlike G4, where the
heuristic was fast and geometrically bounded, the failure in G5 appears to be
computationally bounded by the density of the state space. The algorithm
likely exhausts its ability to efficiently prune the decision tree amidst the
massive number of valid permutations, leading to a computational standstill
despite the optimized memory management.
\end{itemize}

\medskip

\subsection{What we have not done}

The analytic form of $f_{\Gamma}\left(  z\right)  $ can only be deduced from
numerical fits so that the analysis of the mapping $f_{\Gamma}\left(
z\right)  $ depends on the class of functions used in the fit. This fact
prevents a rigorous proof of the identification of the transition from good to
bad alphabets with the transition from regular to chaotic behavior in discrete
mapping. In fact, the existing literature (see, for instance, \cite{15a0}
\cite{15a} \cite{15d} \cite{15e} \cite{15a} \cite{15d} \cite{15e} and
references therein) supports this idea.

Using the heuristics described here above together with some inspiration from
the games of chess and go, one can devise an efficient algorithm to tile large
$N\times N$ squares. Initially, the sacrifice of efficiency is needed and we
have to work by brute force to establish whether or not a given alphabet
$\Gamma$ is good or bad (arriving at $n_{\max}$ steps with $n_{\max}\gg q$).
If $\Gamma$ is bad, there is nothing to do. If $\Gamma$ is good, starting from
the step $n_{\max}$, one changes strategy from brute force to a strategy
adapted to the fact that $\Gamma$ is good (which basically requires to
complete the tiling of a $\left(  n+1\right)  \times\left(  n+1\right)  $
square starting from a seed solution at the previous step). Our data indicate
that, in the good subspace, this strategy works in a polynomial time (indeed,
we have complete tiling of extremely large squares with good $\Gamma s$ which
would be impossible with other algorithms). There is however a severe
technical obstruction which prevents one from proving that the algorithm
always run in a polynomial time in the good subspaces. Namely, there is no
mathematical characterization of the good alphabets (we remind that our
criteria are simply heuristics). Without such a mathematical characterization
it looks unlikely to be able to prove rigorously that the present algorithm
always runs in polynomial time in the good subspace. On the other hand, the
present results are encouraging.

As far as the bad subspace is concerned (although we do not have a rigorous
proof for the reasons explained here above), it is characterized by chaotic
mappings when going from step $n$ to step $n+1$. This implies that the
likelihood to find efficient tilings algorithms in the bad subspace is similar
to the likelihood to find efficient algorithms to predict the asymptotic
behavior of chaotic discrete dynamical system. In other words, in order to
construct tilings of large squares for an alphabet $\Gamma$ in the bad
subspace one must, at the very least, be able to predict the behavior of the
mapping associated to $\Gamma$ after many steps. It seems very unlikely to be
able to construct polynomial time algorithms for the analysis of chaotic mapping.

Despite the lack of rigor, the above considerations suggest that the space of
alphabets is divided in (at least) two regions: one region where polynomial
time algorithms are available and another where polynomial time algorithms are
not available. The fact that the WTP is NP complete together with the
Cook-Levine theorem would then imply that the same is true for all the NP
complete problems and that, consequently, the P vs NP problem has a finer
structure related to the transition from regular to chaotic dynamics.

We will come back on these interesting issues in a future publication.

\section{Conclusions and perspectives}

In the present manuscript, we have constructed an efficient algorithm for the
Wang tiling problem on $N \times N$ squares (where $N \gg q$). This
construction draws inspiration from the games of chess and Go, where the
\textquotedblleft sacrifice of material\textquotedblright\ at the beginning to
secure a greater benefit later constitutes a rigorous and elegant strategy.

Our results indicate that, for the alphabets where a valid seed was
successfully identified, the algorithm runs in polynomial time after the
initial inefficient phase. The behavior of Family G4 provides the most
striking evidence of this \textit{Tractable Regime}: before reaching its
topological impasse at $N=68$, the heuristic demonstrated \textquotedblleft
hyper-efficiency,\textquotedblright\ reducing computational times to near-zero
levels. This explicitly confirms that, once the correct topological channel is
found, the tiling process can indeed be linearized, validating our central hypothesis.

However, the analysis of families G3, G4, and G5 reveals a more complex
scenario. While the failure to reach the asymptotic target $N=1000$ is
proximately caused by the limited set of seeds harvested (due to hardware
constraints on $n_{\max}$), the specific manner in which these available seeds
fail points to distinct structural intricacies. The rarity of viable seeds in
these families implies that the \textit{intrinsic cost} $C(\Gamma)$ to enter
the polynomial regime is significantly higher than in the tractable cases.
Furthermore, the observed failure modes the \textit{topological rigidity} of
G3, the \textit{geometric delayed impasse} of G4, and the \textit{entropic
saturation} of G5, suggest that even if more seeds were available, the
solution space is fraught with local minima and deceptive paths.

These findings suggest that families exhibiting such high initial costs and
complex failure dynamics likely reside in the intermediate \textquotedblleft
Edge of Chaos\textquotedblright\ region. Unlike the tractable alphabets where
valid seeds are abundant and robust, or the chaotic ones where they are
nonexistent, this boundary region is characterized by a solution space where
valid trajectories are mathematically possible but statistically improbable to
discover without massive computational resources. From the dynamical system
viewpoint, this region plays the role of the bifurcations cascade before the
chaotic region.

Consequently, the present results suggest the following picture: P vs NP
problems may possess a \textit{finer structure}. The configuration space of
NP-complete problems can be divided into (at least) three regions. In the
\textit{Tractable Regime}, polynomial-time algorithms are available and
accessible. In the \textit{Chaotic Regime}, efficient algorithms are unlikely
due to computational irreducibility. Finally, our analysis points to the
existence of a complex \textit{Boundary Region} (Edge of Chaos), where the
solvability is undecidable within standard computational limits due to the
extreme scarcity and fragility of the valid solution paths. We hope to address
the formal characterization of this third region in a future publication.

\subsection*{Acknowledgements}

This work has been funded by Fondecyt grants No. 1240048, 1240043 and 1240247
as well as by Proyecto de Exploracion 13250014\textbf{. }The Centro de
Estudios Cientificos (CECs) is funded by the Chilean Government through the
Centers of Excellence Base Financing Program of Conicyt.


\begin{thebibliography}{99}                                                                                               %


\bibitem {PvsNP1}S. A. Cook, "The complexity of theorem-proving procedures",
STOC '71: Proceedings of the third annual ACM symposium on Theory of computing
Pages 151 - 15.

\bibitem {PvsNP2}L. A. Levin, (in Russian) Problems of Information
Transmission, https://www.mathnet.ru/php/archive.phtml?wshow=paper\&jrnid=ppi\&paperid=914\&option\_lang=rus.

\bibitem {PvsNP3}L. Fortnow, "The status of the P versus NP problem",
Communications of the ACM, Volume 52, Issue 9 Pages 78 - 86, https://doi.org/10.1145/1562164.1562186.

\bibitem {PvsNP5}M. Sipser, "Introduction to the Theory of Computation",
Second Edition, International Edition (Thomson Course Technology, 2006).

\bibitem {PvsNP3.5}NSA (2012). "Letters from John Nash" (PDF). Archived (PDF)
from the original on 9 November 2018.

\bibitem {PvsNP4}J. Hartmanis, "Gödel, von Neumann, and the P = NP problem",
Bulletin of the European Association for Theoretical Computer Science. 38: 101--107.



\bibitem {Kirkpatrick1983}S. Kirkpatrick, C. D. Gelatt, and M. P. Vecchi,
"Optimization by Simulated Annealing", Science, Volume 220, Issue 4598, Pages
671--680 (1983). https://doi.org/10.1126/science.220.4598.671



\bibitem {Monasson1999}R. Monasson, R. Zecchina, S. Kirkpatrick, B. Selman,
and L. Troyansky, "Determining computational complexity from characteristic
'phase transitions'", Nature, Volume 400, Pages 133--137 (1999). https://doi.org/10.1038/22055



\bibitem {Mezard2002}M. Mézard, G. Parisi, and R. Zecchina, "Analytic and
algorithmic solution of random satisfiability problems", Science, Volume 297,
Issue 5582, Pages 812--815 (2002). https://doi.org/10.1126/science.1073287



\bibitem {Schuetz2022}M. J. A. Schuetz, J. K. Brubaker, and H. G. Katzgraber,
"Combinatorial optimization with physics-inspired graph neural networks",
Nature Machine Intelligence, Volume 4, Pages 367--377 (2022). https://doi.org/10.1038/s42256-022-00468-6



\bibitem {Gamarnik2021}D. Gamarnik, "The overlap gap property: A topological
barrier to optimizing over random structures", Proceedings of the National
Academy of Sciences (PNAS), Volume 118, Issue 41 (2021). https://doi.org/10.1073/pnas.2108492118



\bibitem {Zdeborova2020}L. Zdeborová, "Understanding deep learning via
statistical physics", Nature Physics, Volume 16, Pages 602--610 (2020). https://doi.org/10.1038/s41567-020-0929-2



\bibitem {Karp1972}R. M. Karp, "Reducibility among Combinatorial Problems",
In: R. E. Miller, J. W. Thatcher (eds) Complexity of Computer Computations,
Plenum Press, New York, Pages 85--103 (1972).



\bibitem {GareyJohnson1979}M. R. Garey and D. S. Johnson, "Computers and
Intractability: A Guide to the Theory of NP-Completeness", W. H. Freeman and
Company, New York (1979).



\bibitem {Preskill2018}J. Preskill, "Quantum Computing in the NISQ era and
beyond", Quantum, Volume 2, Page 79 (2018). https://doi.org/10.22331/q-2018-08-06-79



\bibitem {Wolfram1984}S. Wolfram, "Universality and complexity in cellular
automata", Physica D: Nonlinear Phenomena, Volume 10, Issues 1-2, Pages 1-35
(1984). https://doi.org/10.1016/0167-2789(84)90245-8

\bibitem {[1]}H. Wang, Proving theorems by pattern recognition. II, Bell Syst.
Tech. J. 40 (1961) 1--42.

\bibitem {[2]}R. Berger, The Undecidability of the Domino Problem, Mem. Amer.
Math. Soc., vol.66, 1966.

\bibitem {UndecidPvsNP}A. Klingler, M. van der Eyden, S. Stengele, T.
Reinhart, G. de las Cuevas, \textit{SciPost Phys}. \textbf{14}, 173 (2023).

\bibitem {[15]}S. N. Elaydi, Discrete Chaos: With Applications in Science and
Engineering, Chapman, Hall (CRC Press 2007).

\bibitem {[3]}R. Robinson, Undecidability and nonperiodicity of tilings in the
plane, Invent. Math. 12 (1971) 177--209.

\bibitem {[4]}E. Jeandel, M. Rao, An Aperiodic Set of 11 Wang Tiles, Advances
in Combinatorics, 2021:1, 37 pp.

\bibitem {[4.1]}S. Labbé., A Self-Similar Aperiodic Set of 19 Wang Tiles,
Geometriae Dedicata 201 (2019), pp. 81--109.

\bibitem {[4.2]}J. Kari, A Small Aperiodic Set of Wang Tiles, Discrete
Mathematics 160 (1996), pp. 259--264.

\bibitem {[5]}D. Shechtman, I. Blech, D. Gratias, J. Cahn, Metallic Phase With
Long-Range Orientational Symmetry and No Translational
Symmetry\textquotedblright, Phys. Rev. Lett. 53, 1951--1953 (1984).

\bibitem {[6]}L. Bindi, N. Yao, C. Lin, L. Hollister, C. Andronicos, V.
Distler, M. Eddy, A. Kostin, V. Kryachko, G. MacPherson, W. Steinhardt, M.
Yudovskaya, P. Steinhardt, Natural Quasicrystal With Decagonal Symmetry,
Scientific Reports 5, 9111 (2015).

\bibitem {[7]}D. Levine, P. Steinhardt, Quasicrystals: A New Class of Ordered
Structures, Phys. Rev. Lett. 53 (26 1984), pp. 2477--2480.

\bibitem {[8.1]}J. Mikekisz, A Microscopic Model With Quasicrystalline
Properties, Journal of Statistical Physics 58.5--6 (1990), pp. 1137--1149;
Stable Quasicrystalline Ground States, Journal of Statistical Physics 88.3--4
(1997), pp. 691--711; An Ultimate Frustration in Classical Lattice-Gas Models,
Journal of Statistical Physics 90.1--2 (1998), pp. 285--300.

\bibitem {[9]}A. van Enter, W. Ruszel, Chaotic Temperature Dependence at Zero
Temperature, Journal of Statistical Physics 127.3 (2007), pp. 567--573. 10.1007/s10955-006-9260-2.

\bibitem {[10]}Jean-René Chazottes, M. Hochman, On the Zero-Temperature Limit
of Gibbs States, Communications in Mathematical Physics 297.1 (2010), pp.
265--281. 10.1007/s00220-010-0997-8.

\bibitem {[11]}Leo Gayral. Complexity and Robustness of Tilings with Random
Perturbations, PhD thesis (Université Paul Sabatier - Toulouse III, 2023) https://theses.hal.science/tel-04288597/.

\bibitem {[11.1]}Á. Perales-Eceiza, T. Cubitt, M. Gu, D. Pérez-García, M. M.
Wolf, Undecidability in Physics: a Review, https://doi.org/10.48550/arXiv.2410.16532.

\bibitem {[13]}T. S. Cubitt, D. Perez-Garcia, M. M. Wolf, M.M., Undecidability
of the spectral gap. Nature 528, 207 (2015); Comment on \textquotedblleft on
the uncomputability of the spectral gap\textquotedblright. arXiv:1603.00825.

\bibitem {[14]}N. Shiraishi, K. Matsumoto, Undecidability in quantum
thermalization, Nature Communications 12, 5084 (2021).

\bibitem {15a0}G. J. Chaitin, \textit{Scientific American} \textbf{232} (5),
47 (1975); \textit{Int. I. Theor, Phys} \textbf{22}, 941 (1982).

\bibitem {15a}M. O. Rabin, \textit{Trans. Amer. Math. Soc.} \textbf{141}
(1969), 1-35.

\bibitem {15d}C. Agnes, M. Rasetti, \textit{Nuovo Cimento} \textbf{106}, 879
(1991); \textit{Chaos, Solitons \& Fractals} \textbf{5}, 161-175 (1995).

\bibitem {15e}M. Rasetti, \textit{Chaos, Solitons \& Fractals} \textbf{5},
133-138 (1995).

\bibitem {[16]}Burn, D. H. and Elnur, M. A. H. (2002), Detection of hydrologic
trends and variability. Journal of Hydrology, 255(1-4), 107-122.

\bibitem {[17]}Croux, C., and Dehon, C. (2010), Influence functions of the
Spearman and Kendall correlation measures. Statistical Methods, Applications,
19(4), 497-515.

\bibitem {[18]}Hernandez, W., Mendez, A., Zalakeviciute, R., and Diaz-Marquez,
A., M., (2020). Analysis of the Information Obtained from PM2.5 Concentration
Measurements in an Urban Park, https://doi.org/10.1109/TIM.2020.2966360

\bibitem {[19]}Grünbaum, B., and Shephard, G. C. (1987). Tilings and patterns.
W. H. Freeman and Company.



\bibitem {Arora2009}S. Arora and B. Barak, "Computational Complexity: A Modern
Approach", Cambridge University Press (2009).



\bibitem {Pak2013}I. Pak and J. Yang, "Tiling simply connected regions with
rectangles", Journal of Combinatorial Theory, Series A, Volume 120, Issue 7,
Pages 1804--1816 (2013). https://doi.org/10.1016/j.jcta.2013.06.008



\bibitem {Aamand2023}A. Aamand, M. Abrahamsen, T. D. Ahle, and P. M. R.
Rasmussen, "Tiling with Squares and Packing Dominos in Polynomial Time", ACM
Transactions on Algorithms, Volume 19, Issue 3, Article 30, Pages 1--28
(2023). https://doi.org/10.1145/3597932



\bibitem {Smith2024}D. Smith, J. S. Myers, C. S. Kaplan, and C.
Goodman-Strauss, "An aperiodic monotile", Combinatorial Theory, Volume 4,
Issue 1 (2024). https://doi.org/10.5070/C64163839

\bibitem {MarcoFabrizio}F. Canfora, M. Cede$\widetilde{n}$o, \textit{Partial
decidability protocol for the Wang tiling problem from statistical mechanics
and chaotic mapping}, \textit{Physica} \textbf{A 682} (2026) 131187. DOI: 10.1016/j.physa.2025.131187.

\bibitem {[23]}Karel Culik, \textit{An aperiodic set of 13 Wang tiles}.
Discrete Mathematics 160 (1996) 245-251.
\end{thebibliography}
\end{document}